\documentclass[aps,pra,amsfonts,amssymb,twocolumn,amsmath,preprintnumbers,nofootinbib,floatfix,
showpacs,superscriptaddress]{revtex4-1}%
\usepackage[dvips]{graphics}
\usepackage{graphicx}
\usepackage{dcolumn}
\usepackage{bm}
\usepackage{amsmath}
\usepackage{amsfonts}
\usepackage{amssymb}
\usepackage{xcolor}
\usepackage{subfigure}
\usepackage{hyperref,hypcap}
\usepackage{braket}
\usepackage{commath}%
\providecommand{\U}[1]{\protect\rule{.1in}{.1in}}
\begin{document}

\title{Theory of phonon side jump contribution in Anomalous Hall transport}

\author{Cong Xiao}
\email{congxiao@utexas.edu}
\affiliation{Department of Physics, The University of Texas at Austin, Austin, Texas 78712, USA}

\author{Ying Liu}
%\email{ying$_$liu@mymail.sutd.edu.sg}
\affiliation{Research Laboratory for Quantum Materials, Singapore University of Technology and Design, Singapore 487372, Singapore}

\author{Ming Xie}
\affiliation{Department of Physics, The University of Texas at Austin, Austin, Texas 78712, USA}

\author{Shengyuan A. Yang}
\affiliation{Research Laboratory for Quantum Materials, Singapore University of Technology and Design, Singapore 487372, Singapore}

\author{Qian Niu}
\affiliation{Department of Physics, The University of Texas at Austin, Austin, Texas 78712, USA}

\begin{abstract}
The role of electron-phonon scattering in finite-temperature anomalous Hall
effect is still poorly understood. In this work, we present a Boltzmann theory
for the side-jump contribution from electron-phonon scattering, which is
derived from the microscopic quantum mechanical theory. We show that the resulting phonon side-jump conductivity
generally approaches different limiting values in the high and low temperature limits, and
hence can exhibit strong temperature dependence in the intermediate temperature regime. Our
theory is amenable to $ab$ $initio$ treatment, which makes
quantitative comparison between theoretical and experimental results possible.
\end{abstract}
\maketitle

\section{Introduction}
Electron-phonon scattering plays a key role in electronic transport in
crystalline solids \cite{Ziman1960,White1958}. For longitudinal transport,
electron-phonon scattering limits the intrinsic mobility, and its effect can
now be well evaluated via a combination of the first-principles band structure
calculation and semiclassical Boltzmann approach
\cite{AllenLOVA,Li2015,Louie2016,GaAs,MoS2,Kim2010,Mauri2014,Giustino2017}.
However, its role in the anomalous Hall transport is much more subtle
\cite{Tian2009,Nagaosa2010,Shitade2012,Ebert2013,Hou2015,Otani2019,Meng2016,Xia2017},
and a clear understanding has yet to be achieved.

Theoretical study of the anomalous Hall transport has been mostly performed
with \emph{static} impurities \cite{Nagaosa2010}. In the weak scattering
regime, anomalous Hall conductivity is known to have three important
contributions arising from different mechanisms in the semiclassical picture
\cite{Sinitsyn2008,Sinitsyn2007}: intrinsic contribution from Berry curvatures in band
structures \cite{Jungwirth2002,Yao2004}, side jump from electron coordinate shift during
scattering \cite{Berger1970,Sinitsyn2006}, and skew scattering from the
asymmetric part of the scattering rate \cite{Smit,Borunda2007}.
Particularly, side jump is a very peculiar contribution in that although it
results from scattering, its value is found to be independent of the impurity
concentration for static impurity scattering \cite{Berger1970,Nagaosa2010,Kovalev2010,Freimuth2011}.

Will phonon scattering be any different? Typically, the phonon energy scale
($k_{B}T$) is much less than the Fermi energy $\epsilon_{F}$, so the energy
transfer in phonon scattering would be negligible. It seems that the phonon
side jump contribution should be similar to that of static impurities, and
hence it should be insensitive to temperature ($T$) \cite{Berger1970,Bruno2001,Dugaev2001}. This speculation
has gained support
from experiments performed at elevated temperatures where the longitudinal resistivity shows
linear in $T$ dependence \cite{Dheer1967,Coleman1974,Miyasato2007}. Recently, researchers
do realize that the side-jump from phonon and impurity scattering can be
different, thereby the change of their relative importance with temperature
can lead to $T$-dependent behavior \cite{Yang2011,Hou2015}. However, the
$T$-independence of the phonon side-jump contribution alone has not been doubted.

In a very recent work \cite{Xiao2019}, it is realized that the phonon side jump
contribution can indeed be $T$-dependent. The key ingredient is the $T$-dependent
phonon occupation number, which makes the average momentum transfer, i.e., the
effective range, of electron-phonon scattering $T$-dependent. By analogy with
the recently revealed sensitivity of the anomalous Hall conductivity to the
scattering range of static random impurities \cite{Ado2017}, one can
understand qualitatively the $T$-dependence of phonon side jump.

However, we do not yet have a theory of phonon side jump with quantitative
predictive power, accounting for the dynamical and inelastic nature of
electron-phonon scattering. Here, we develop such a theory within the
semiclassical Boltzmann framework. Surely, one may choose to construct a
theory on a more fundamental level, with a fully quantum field theoretical
treatment, and there were indeed a few attempts in the past
\cite{Leribaux1966,Lyo1973}. Unfortunately, due to the complexity in modeling phonon
scattering, such transport theories are extremely complicated, lacking
physical transparency, and too difficult to be combined with $ab$ $initio$
calculations for real materials. In comparison, the semiclassical theory
presented here enjoys the advantages of being physically intuitive and
easily implementable with $ab$ $initio$ calculations. As an application of
this theory, we show that the phonon side jump conductivity generally saturates to two different values in low
and high temperature limits, and the strong $T$-dependence naturally appears in
the temperature regime in-between.

Our paper is organized as follows. In Sec.~II we review the semiclassical theory for side jump from impurity scattering, and propose the new theory for phonon-induced side-jump in a heuristic way. In Sec.~III, we present a general argument for the $T$-dependence of phonon side jump conductivity. This $T$-dependence is explicitly demonstrated in Sec.~IV, by applying our theory to study the concrete massive Dirac model. Finally, in Sec.~V, we discuss the possible experimental scheme to confirm our result and conclude this work. The detailed derivation of our theory is presented in the Appendix.

\section{Boltzmann theory for phonon side jump}
We start by reviewing the theory for side jump induced by impurity scattering.
The semiclassical nonequilibrium distribution function $f$ for electron
wave-packets in phase space is governed by the Boltzmann equation:
\begin{equation}
\left(  \partial_{t}+{\dot{\bm r}}\cdot\partial_{\bm{r}}+{\dot{\bm k}}%
\cdot\partial_{\bm{k}}\right)  f=I_{\text{coll}}\left[  f\right]  .
\end{equation}
With a uniform dc electric field and in the steady state, the linearized
equation takes the form of (set $e=\hbar=1$)
\begin{subequations}
\begin{equation}
\bm{E}\cdot\bm{v}_{\ell}^{0}\partial_{\epsilon_{\ell}}f_{\ell}^{0}=-\sum
_{\ell^{\prime}}\big[w_{\ell^{\prime}\ell}f_{\ell}\left(  1-f_{\ell^{\prime}%
}\right)  -\left(  \ell\leftrightarrow\ell^{\prime}\right)  \big],
\label{SBE-conventional}%
\end{equation}
where the added subscript $\ell\equiv n\bm k$ labels the Bloch state,
$\bm{v}_{\ell}^{0}=\partial_{\bm k}\epsilon_{\ell}$ is the band velocity,
$f^{0}$ is the equilibrium Fermi-Dirac distribution function, the collision
term $I_{\text{coll}}\left[  f\right]  $ on the right hand side is explicitly
written out with a scattering-out term ($\ell\rightarrow\ell^{\prime}$) and a
scattering-in term ($\ell^{\prime}\rightarrow\ell$), and $w$ is the
corresponding scattering rate. We may write $f_{\ell}=f_{\ell}^{0}+\delta
f_{\ell}=f_{\ell}^{0}+(-\partial_{\epsilon_{\ell}}f_{\ell}^{0})g_{\ell}$,
where the second equality indicates the fact that the nonequilibrium deviation
should be around the Fermi surface and $g_{\ell}$ is a smooth function of
energy and momentum. In the \emph{absence} of side jump, using the principle
of detailed balance, namely, $w_{\ell^{\prime}\ell}f_{\ell}^{0}(1-f_{\ell
^{\prime}}^{0})=w_{\ell\ell^{\prime}}f_{\ell^{\prime}}^{0}(1-f_{\ell}^{0})$,
and keeping terms to linear order in $E$, one can show that
Eq.~(\ref{SBE-conventional}) can be put into the following form for $g_{\ell}%
$:
\begin{equation}
\bm{E}\cdot\bm{v}_{\ell}^{0}=\sum_{\ell^{\prime}}\frac{1-f_{\ell^{\prime}}%
^{0}}{1-f_{\ell}^{0}}w_{\ell^{\prime}\ell}\left(  {g}_{\ell}-{g}_{\ell
^{\prime}}\right)  . \label{SBE-n}%
\end{equation}
Here, we emphasize that Eqs.~(\ref{SBE-conventional}) and (\ref{SBE-n}) are
valid for both static (impurity) and dynamical (phonon) disorder. For static
impurities, the factor $(1-f_{\ell^{\prime}}^{0})/(1-f_{\ell}^{0})$ in
Eq.~(\ref{SBE-n}) (which may be called the Pauli factor) becomes unity, and
the result reduces to the familiar one in textbooks.

Side jump refers to the coordinate shift of the electron wave-packet during
scattering, for which Sinitsyn \emph{et al.} have derived a general expression
\cite{Sinitsyn2006}:
\end{subequations}
\begin{equation}
\delta\bm{r}_{\ell^{\prime}\ell}=-\delta\bm{r}_{\ell\ell^{\prime}%
}=\bm{A}_{\ell^{\prime}}-\bm{A}_{\ell}-\left(  \partial_{\bm{k}}%
+\partial_{\bm{k}^{\prime}}\right)  \arg V_{\ell^{\prime}\ell}, \label{shift}%
\end{equation}
where $\bm{A}_{\ell}=i\langle u_{\ell}|\partial_{\bm{k}}|u_{\ell}\rangle$ is
the Berry connection, $|u\rangle$ is the periodic part of the Bloch state, and
$V_{\ell^{\prime}\ell}$ is the scattering matrix element.

Due to this coordinate shift, the $\bm{E}$ field does a nonzero work in
scattering, which has to be accounted for in energy conservation
\cite{Sinitsyn2008}. For static impurity scattering, one then has
$\epsilon_{\ell^{\prime}}=\epsilon_{\ell}+\bm E\cdot\delta\bm r_{\ell^{\prime
}\ell}$. Consequently, the equilibrium distribution no longer annihilate the
collision term, because $f_{\ell}^{0}-f_{\ell^{\prime}}^{0}\approx
-\partial_{\epsilon_{\ell}}f_{\ell}^{0}\bm{E}\cdot\delta\bm{r}_{\ell^{\prime
}\ell}$, and from the Boltzmann equation, this leads to an additional
(anomalous) correction to the distribution function: $\delta f_{\ell}%
^{a}=\left(  -\partial_{\epsilon_{\ell}}f_{\ell}^{0}\right)  {g}_{\ell}^{a}$,
satisfying
\begin{equation}
\bm{E}\cdot\sum_{\ell^{\prime}}w_{\ell^{\prime}\ell}\delta\bm{r}_{\ell
^{\prime}\ell}=-\sum_{\ell^{\prime}}w_{\ell^{\prime}\ell}\left(  {g}_{\ell
}^{a}-{g}_{\ell^{\prime}}^{a}\right)  . \label{SBE-a-static}%
\end{equation}
Thus, the out-of-equilibrium part of the distribution is
\begin{equation}
\delta f_{\ell}=\delta f_{\ell}^{n}+\delta f_{\ell}^{a}=\left(  -\partial
_{\epsilon_{\ell}}f_{\ell}^{0}\right)  \left(  {g}_{\ell}^{n}+{g}_{\ell}%
^{a}\right)  ,
\end{equation}
where the terms with superscript $n$ refer to the \textquotedblleft normal"
contribution, satisfying Eq.~(\ref{SBE-n}) without the side jump effect.
Meanwhile, the side jump also corrects the electron velocity, which becomes
\begin{equation}
\bm v_{\ell}=\bm v_{\ell}^{0}+\bm v_{\ell}^{\text{bc}}+\bm v_{\ell}%
^{\text{sj}}.
\end{equation}
Here, $\bm v_{\ell}^{\text{bc}}=\bm\Omega_{\ell}\times\bm E$ is the anomalous
velocity induced by Berry curvature $\bm\Omega_{\ell}=\partial_{\bm k}%
\times\bm A_{\ell}$, and
\begin{equation}
\bm v_{\ell}^{\text{sj}}=\sum_{\ell^{\prime}}w_{\ell^{\prime}\ell}%
\delta\bm r_{\ell^{\prime}\ell} \label{sjv}%
\end{equation}
is called the side jump velocity. Applying the $\bm{E}$ field in the $x$
direction, then the intrinsic anomalous Hall current is given by
$j_{\text{AH}}^{\text{in}}=\sum_{\ell}f_{\ell}^{0}(\bm v_{\ell}^{\text{bc}%
})_{y}$. The side jump induced Hall current, which is the focus of this paper,
contains two terms to linear order in $E$:
\begin{equation}
j_{\text{AH}}^{\text{sj}}=j_{\text{AH}}^{\text{sj}(1)}+j_{\text{AH}%
}^{\text{sj}(2)}=\sum_{\ell}\delta f_{\ell}^{n}(\bm v_{\ell}^{\text{sj}}%
)_{y}+\sum_{\ell}\delta f_{\ell}^{a}(\bm v_{\ell}^{0})_{y}. \label{sjj}%
\end{equation}
Note that counting the order in relaxation time $\tau$, $\delta f^{n}\sim\tau
$, $\delta f^{a}\sim\tau^{0}$, $v^{0}\sim\tau^{0}$, and $v^{\text{sj}}\sim
\tau^{-1}$, so both terms in $j_{\text{AH}}^{\text{sj}}$ are on the order of
$\tau^{0}$. For static impurities, the side jump contribution is independent
of the impurity density as well as the scattering potential strength. The
above semiclassical theory was shown to be consistent with fully quantum
mechanical treatment for static impurities \cite{Sinitsyn2007}. Particularly, the
side jump velocity in Eq.~(\ref{sjv}) was found to correspond to the
scattering-induced band-off-diagonal elements of the \textit{out-of-equilibrium}
density matrix \cite{Sinitsyn2006,KL1957,Xiao2018KL}.

Now let's turn to phonon scattering. In the following, we present a heuristic
argument for the theory. First of all, we note that
Eqs.~(\ref{SBE-conventional}) and (\ref{shift}) apply for dynamical disorder
like phonons as well. Like before, the side jump leads to an additional work
done by the $\bm{E}$ field, modifying the relation between $\epsilon_{\ell}$
and $\epsilon_{\ell^{\prime}}$, with
\begin{equation}
\tilde{\epsilon}_{\ell^{\prime}}=\epsilon_{\ell}+\bm E\cdot\delta
\bm r_{\ell^{\prime}\ell}\pm\omega_{q},
\end{equation}
where the last term indicates the absorption or emission of a phonon with mode
label $q$. Then the linearized Boltzmann equation becomes (details in Appendix A)
\begin{equation}
\bm{E}\cdot\bm{v}_{\ell}^{0}=\sum_{\ell^{\prime}}\frac{1-f^{0}\left(
\epsilon_{\ell^{\prime}}\right)  }{1-f^{0}\left(  \epsilon_{\ell}\right)
}w_{\ell^{\prime}\ell}\left(  {g}_{\ell}-{g}_{\ell^{\prime}}+\bm{E}\cdot
\delta\bm{r}_{\ell^{\prime}\ell}\right)  . \label{SBE-new}%
\end{equation}
where $\epsilon_{\ell^{\prime}}=\epsilon_{\ell}\pm\omega_{q}$.
Subtracting Eq.~(\ref{SBE-n})
from Eq.~(\ref{SBE-new}) shows that the anomalous correction to the
distribution due to side jump satisfies the equation
\begin{equation}
\bm{E}\cdot\sum_{\ell^{\prime}}\frac{1-f_{\ell^{\prime}}^{0}}{1-f_{\ell}^{0}%
}w_{\ell^{\prime}\ell}\delta\bm{r}_{\ell^{\prime}\ell}=-\sum_{\ell^{\prime}%
}\frac{1-f_{\ell^{\prime}}^{0}}{1-f_{\ell}^{0}}w_{\ell^{\prime}\ell}\left(
{g}_{\ell}^{a}-{g}_{\ell^{\prime}}^{a}\right)  . \label{SBE-a}%
\end{equation}
Comparing Eq.~(\ref{SBE-a}) with Eq.~(\ref{SBE-a-static}) \textit{suggests}
that the proper definition for the phonon side jump velocity should be
\begin{equation}
\bm{v}_{\ell}^{\text{sj}}=\sum_{\ell^{\prime}}\frac{1-f_{\ell^{\prime}}^{0}%
}{1-f_{\ell}^{0}}w_{\ell^{\prime}\ell}\delta\bm{r}_{\ell^{\prime}\ell}.
\label{sj velocity}%
\end{equation}
The above three equations are the main results of this paper.
Here, the main difference between Eqs.~(\ref{SBE-a},\ref{sj velocity}) and
Eqs.~(\ref{SBE-a-static},\ref{sjv}) is the appearance of the Pauli factor,
which, as we have discussed before, reflects the dynamical character of phonon
scattering. For static impurity scattering, the Pauli factor becomes unity,
and the theory correctly recovers the familiar one. In metals the Pauli factor
is important for acoustic phonons in the low-$T$ regime where $\omega_{q}$ is
of the order of $k_{B}T$, thus the electronic occupancy $f_{\ell}^{0}$ and
$f_{\ell^{\prime}}^{0}$ differ significantly. Whereas in semiconductor
low-dimensional electron systems with small Fermi energy, the Pauli factor is
also important for highly inelastic optical phonons \cite{Mauri2014}.

With the new definition of the side jump velocity in Eq.~(\ref{sj velocity})
and with $g_{\ell}^{a}$ solved from Eq.~(\ref{SBE-a}), the side jump current
will still be calculated with Eq.~(\ref{sjj}). This completes our
semiclassical theory for phonon side jump.

This theory, albeit seemingly simple and intuitive, is in fact nontrivial. Its
justification requires tedious derivation from microscopic theories of
coupled electron-phonon system. We have demonstrated that the theory can be derived from
two different fundamental approaches: the density matrix equation of motion
approach~\cite{Argyres1961} and the Lyo-Holstein's transport
theory~\cite{Lyo1973,Holstein1964}. The details are relegated to Appendices C and D.

\section{Temperature dependence of phonon side jump}

As we have mentioned at the beginning, for $k_{B}T\ll\epsilon_{F}$, the common
belief is that the phonon side jump Hall conductivity $\sigma_{\text{AH}%
}^{\text{sj}}$ $(\equiv j_{\text{AH}}^{\text{sj}}/E_{x})$ should be
independent of the strength of disorder scattering (so its value remains the
same even if the disorder density approaches zero), and hence it should have
little $T$ dependence. As an application of our theory, we shall see that this naive conclusion is generally
incorrect in the case where side jump arises from spin-orbit-coupled Bloch
electrons scattered off phonons.

Consider the low-$T$ limit, which is specified by $T\ll T_{D}$, where $T_{D}$
is the Debye temperature (Note that in this discussion, $\epsilon_{F}$ is
always assumed to be the largest energy scale). For such case, the scattering
is dominated by long wavelength acoustic phonons, which is short ranged in
momentum space. Hence, the coordinate shift reduces to $\delta\bm{r}_{\ell
^{\prime}\ell}\approx\mathbf{\Omega}_{\ell}\times\left(  \bm{k}^{\prime
}-\bm{k}\right)  $. From Eq.~(\ref{SBE-a}), we find that $g_{\ell}%
^{a}=\bm{E}\cdot\left(  \mathbf{\Omega}_{\ell}\times\bm{k}\right)  $, whose
contribution to the Hall conductivity (corresponding to $j_{\text{AH}%
}^{\text{sj}(2)}$) is $\sigma_{\text{AH}}^{\text{sj}(2)}=-\sum_{\ell}\left(
\mathbf{\Omega}_{\ell}\times\bm{k}\right)  _{x}\partial_{k_{y}}f_{\ell}^{0}$.
Meanwhile, straightforward calculation of $j_{\text{AH}}^{\text{sj}(1)}%
$ yields $\sigma_{\text{AH}}^{\text{sj}(1)}=\sum_{\ell}\left(
\mathbf{\Omega}_{\ell}\times\bm{k}\right)  _{y}\partial_{k_{x}}f_{\ell}^{0}$.
Thus, the phonon side jump Hall conductivity in the low-$T$ limit can be put
into a compact form of
\begin{equation}
\sigma_{\text{AH}}^{\text{sj}}=-\sum_{\ell}\left[  \left(  \mathbf{\Omega
}_{\ell}\times\bm{k}\right)  \times\partial_{\bm{k}}f_{\ell}^{0}\right]  _{z}.
\end{equation}
For two-dimensional systems, the Berry curvature has only $z$-component
$\mathbf{\Omega}_{\ell}=\Omega_{\ell}\bm\hat{z}$, so the above result can be
further simplified as
\begin{equation}
\sigma_{\text{AH}}^{\text{sj}}=\sum_{\ell}\Omega_{\ell}\ \bm{k}\cdot
\partial_{\bm{k}}f_{\ell}^{0}.\label{SJHC-lowT}%
\end{equation}

In the high-$T$ limit with $T\gg T_{D}$, we find that the major $T$ dependence
comes from the scattering rate, which can be approximated as
\begin{equation}
w_{\ell^{\prime}\ell}\approx4\pi\left\vert \langle u_{\ell^{\prime}}|u_{\ell
}\rangle\right\vert ^{2}\left\vert V_{\bm{k}^{\prime}\bm{k}}^{\text{o}%
}\right\vert ^{2}\frac{k_{B}T}{\omega_{q}}\delta\left(  \epsilon_{\ell
}-\epsilon_{\ell^{\prime}}\right)  .
\end{equation}
Here, we have written $V_{\ell^{\prime}\ell}=V_{\bm{k}^{\prime}\bm{k}}%
^{\text{o}}\langle u_{\ell^{\prime}}|u_{\ell}\rangle$, with $V_{\bm{k}^{\prime
}\bm{k}}^{\text{o}}$ the plane-wave part of the electron-phonon scattering
matrix element, and we have used the relation that $N_{q}\simeq(N_{q}+1)\simeq
k_{B}T/\omega_{q}$ in the high-$T$ limit, where $N_{q}$ is the Bose-Einstein
distribution for the phonon mode $q$. Hence in the high-$T$ limit, we have
${g}^{n}\sim T^{-1}$, $v^{\text{sj}}\sim T$, $g^{a}\sim T^{0}$, and thus
$\sigma_{\text{AH}}^{\text{sj}}$ should saturate to a $T$-independent constant
value. Although we cannot write down a compact analytical expression for this
limiting value (because of the complicated model-dependent interband
scattering processes), it is clear that this value should generally be
different from the low-$T$ limit value in Eq.~(\ref{SJHC-lowT}). This analysis
demonstrates that the phonon side jump conductivity $\sigma_{\text{AH}%
}^{\text{sj}}$ approaches different values in the low-$T$ and high-$T$ limits,
therefore pronounced $T$ dependence must exist in the intermediate range when
the two limiting values differ by a significant amount.

\section{Application to Massive Dirac model}
\begin{figure}[tbh]
\includegraphics[width=0.8\columnwidth]{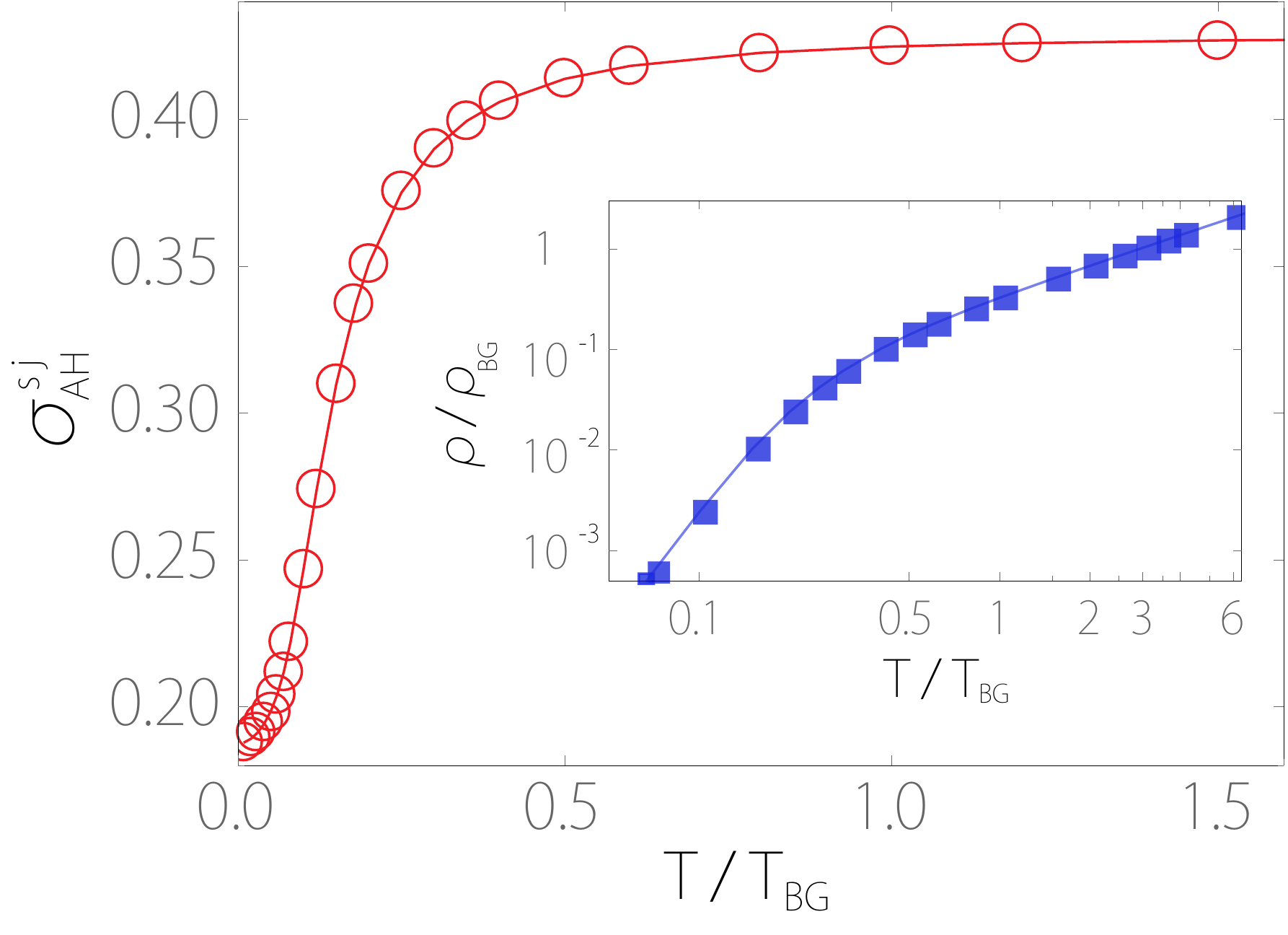}\caption{Temperature-dependence
of the acoustic phonon limited side-jump Hall conductivity (in units of $e^2/h$) in the
 massive Dirac model. Inset: $T$-dependent longitudinal
resistivity $\rho$ shown in the log-log plot. $\rho_\text{BG}$ is a characteristic
resistivity defined from the expression of $\rho$ \cite{Kim2010}, whose
value is not important here. Here, $T_\text{BG}$ is the Bloch-Gruneisen temperature (see
the text), and we have set $\epsilon_F=2\Delta$.}%
\label{fig:T-dependence}%
\end{figure}

In this section, we illustrate the above points by a concrete model calculation using
our theory. We take the two-dimensional massive Dirac model
\begin{equation}
\mathcal{H}_{0}=v\left(  k_{x}{\sigma}_{x}+k_{y}{\sigma}_{y}\right)
+\Delta{\sigma}_{z},
\end{equation}
which is considered as the minimal model for studying anomalous Hall effect.
Here, $v$ and $\Delta$ are model parameters, and the $\sigma$'s are the Pauli
matrices representing the two Dirac bands. Recalling that we work under the
condition $k_{B}T\ll\epsilon_{F}$, hence, to proceed analytically, we neglect
the phonon energy in the scattering, such that $k=k^{\prime}$ for the two
electronic states before and after scattering \cite{Ziman1960}. We consider
the metallic regime $\epsilon_{F}>\Delta$ with low carrier density such that
the Fermi surface is much smaller than the size of Brillouin zone. Thus the
Umklapp process does not occur. We assume the scattering is dominated by acoustic phonons, and the
electron-phonon coupling can be described by the deformation potentials (details in Appendix B). The
coordinate shift for this model can be found as
\begin{equation}
\left(  \delta\bm{r}_{\bm k^{\prime}\bm k}\right)  _{y}=-\frac{\Delta v^{2}%
}{2(\Delta^{2}+\left(  vk\right)  ^{2})^{3/2}}\frac{\left(  k_{x}^{\prime
}-k_{x}\right)  }{\left\vert \langle u_{\bm{k}^{\prime}}|u_{\bm{k}}%
\rangle\right\vert ^{2}}.
\end{equation}
And straightforward calculation (see Appendix B for details) based on our theory leads to
\begin{equation}
\sigma_{\text{AH}}^{\text{sj}}=\frac{1}{4\pi}\frac{\Delta}{\epsilon_{F}%
}\left[  1-\left(  \frac{\Delta}{\epsilon_{F}}\right)  ^{2}\right]
\mathcal{R}(\epsilon_{F},T),
\end{equation}
where the temperature dependence is dumped into the factor $\mathcal{R}$
defined as $\mathcal{R}\equiv\tau^{\text{tr}}/\tau^{\text{sj}}$, where
$\tau^{\text{tr}}$ is the transport relaxation time with
\begin{equation}
(\tau^{\text{tr}})^{-1}=\sum_{\bm{k}^{\prime}}\frac{1-f_{\bm k^{\prime}}^{0}%
}{1-f_{\bm k}^{0}}w_{\bm{k}^{\prime}\bm{k}}\left(  1-\cos\phi_{\bm{kk}^{\prime
}}\right)  ,\label{transport time}%
\end{equation}
$\tau^{\text{sj}}$ is defined as
\begin{equation}
(\tau^{\text{sj}})^{-1}=\sum_{\bm{k}^{\prime}}\frac{1-f_{\bm k^{\prime}}^{0}%
}{1-f_{\bm k}^{0}}\frac{w_{\bm{k}^{\prime}\bm{k}}}{|\langle u_{\bm k^{\prime}%
}|u_{\bm k}\rangle|^{2}}\left(  1-\cos\phi_{\bm{kk}^{\prime}}\right)  ,
\end{equation}
and $\phi_{\bm{kk}^{\prime}}$ is the angle between $\bm k$ and $\bm k^{\prime
}$. In the low-$T$ and high-$T$ limits, we have respectively
\begin{equation}
\mathcal{R}\rightarrow1\text{ \ and \ }\mathcal{R}\rightarrow4[1+3\left(
\Delta/\epsilon_{F}\right)  ^{2}]^{-1}.
\end{equation}
This demonstrates clearly that the phonon side jump contribution
approaches different values in the low-$T$ and high-$T$ limits. This behavior
is illustrated in Fig.~\ref{fig:T-dependence}, where the $T$-dependence
in the intermediate regime is obtained by assuming isotropic Debye spectrum
$\omega_{q}=c_{s}q$ ($c_{s}$ is the sound velocity). The $T$-dependence of the
phonon side-jump contribution becomes apparent when $T<T_\text{BG}/2$. Note that in the same
regime, one can show that the phonon-limited longitudinal resistivity also departs from the linear-$T$
scaling (see the inset of Fig.~\ref{fig:T-dependence}). Here $T_\text{BG}=2\hbar
c_{s}k_{F}/k_{B}$ is the Bloch-Gruneisen temperature, which marks the lower
boundary of the high-$T$ equipartition regime ($\rho\sim T$) in
two-dimensional metallic systems \cite{Kim2010}.

\section{Discussion and conclusion}
We discuss the possible experimental scheme to confirm our result.
The $d$-band ferromagnetic transition metals such as Fe and Co offer suitable platforms, because their band splittings are much larger than room temperature, and the Curie temperatures are much higher than $T_{D}$. It follows that the intrinsic Berry-curvature
contribution to the anomalous Hall conductivity should be $T$-insensitive up to room
temperature. In order to observe the electron-phonon dominated behavior at
lower temperatures (where $\rho$ deviates from the linear-in-$T$ scaling), one needs to work with
high-purity samples (the resistance ratio should be at least
100), which are experimentally accessible \cite{White1958}. The skew
scattering contribution due to non-Gaussian impurity correlations should be
first subtracted from the data. This can be done by using the recently developed
thin-film approach \cite{Hou2015,Yue2017}. In this approach one can limit the
scattering of electrons
to two main sources --- the interface roughness and phonons, and achieve
independent control of each one by tuning
the film thickness and the temperature \cite{Yue2017}. The
aforementioned skew scattering Hall conductivity in this case is given by $\alpha_{0}\rho_{0}/\rho^{2}$,
where $\rho_{0}$ is the residual resistivity, and $\alpha_{0}$ is a
system-specific parameter independent of film thickness that can be
determined by tuning film thickness in the low-$T$ regime \cite{Hou2015}. After
subtracting the skew scattering contribution, one
can verify the $T$-dependence of the side-jump conductivity predicted here. Quantitatively,
one can further subtract the $T$-insensitive intrinsic contribution obtained
from $ab$ $initio$ method \cite{Yao2004}, and then compare the remaining
 to the phonon side-jump Hall conductivity yielded by
the $ab$ $initio$ Boltzmann approach based on our result.

In conclusion, we have proposed a semiclassical Boltzmann theory for the phonon side jump contribution in the anomalous Hall effect. This intuitive theory has been derived from microscopic quantum mechanical transport theories of coupled electron-phonon systems. We demonstrate that the phonon side jump anomalous Hall conductivity can generally be temperature-dependent, which disproves the previous common belief that this contribution is $T$-independent. The possible experimental scheme to confirm our result has been discussed. The proposed Boltzmann formalism can be easily implementable with $ab$ $initio$ calculations, making quantitative comparison between theoretical and experimental results possible.

\begin{acknowledgments}
We thank Yi Liu and Liang Dong for helpful discussions. Q.N. is supported by DOE (DE-FG03-02ER45958, Division of Materials Science and Engineering) on the semiclassical formulation of this work. C.X. is supported by NSF (EFMA-1641101) and Welch Foundation (F-1255). M.X. is supported by the Welch Foundation under grant TBF1473. Y.L. and S.A.Y. are supported by Singapore Ministry of Education AcRF Tier 2 (MOE2017-T2-2-108).
\end{acknowledgments}

C. X. and Y. L. contributed equally to this work.

\appendix

\begin{widetext}
\section{Heuristic argument for the side jump in the Bloch-Boltzmann
equation}

In the presence of a dc weak uniform electric field $\mathbf{E}$ and weak
static disorder, the conventional Boltzmann equation for charge carriers
(charge $e$) in nonequilibrium steady state reads \cite{Ziman1960}%
\begin{equation}
e\mathbf{E}\cdot\mathbf{v}_{\ell}^{0}\left(  -\frac{\partial f_{\ell}^{0}%
}{\partial\epsilon_{\ell}}\right)  =\sum_{\ell^{\prime}}\left(  w_{\ell
^{\prime}\ell}f_{\ell}-w_{\ell\ell^{\prime}}f_{\ell^{\prime}}\right)  .
\end{equation}
In the case of static disorder there is no room \cite{KL1957} for the Pauli
blocking factors $\left(  1-f_{\ell^{\prime}}\right)  $ and $\left(
1-f_{\ell}\right)  $, which were introduced into the collision term of the
Boltzmann equation \textit{phenomenologically} by F. Bloch when studying
phonon-limited mobility in metals in order to ensure the equilibrium Fermi
distribution (rather than Bose or Boltzmann distributions) for $f_{\ell}^{0}$
\cite{Allen1978}. In the case of dynamical disorder such as phonons, the
Bloch-Boltzmann equation takes the form of Eq.~(\ref{SBE-conventional}), where
$w_{\ell^{\prime}\ell}$ and $w_{\ell\ell^{\prime}}$ are calculated in the
quantum mechanical perturbation theory. The collision term
is considered only in the linear response regime. To the lowest order in Born expansion, the principle of microscopic detailed balance holds, as
can be directly verified for electron-phonon scattering. Thus $w_{\ell
\ell^{\prime}}=w_{\ell^{\prime}\ell}e^{\beta\left(  \epsilon_{\ell^{\prime}%
}-\epsilon_{\ell}\right)  }$, and the Bloch-Boltzmann equation reads%
\begin{equation}
e\mathbf{E}\cdot\mathbf{v}_{\ell}^{0}\left(  -\frac{\partial f_{\ell}^{0}%
}{\partial\epsilon_{\ell}}\right)  =\sum_{\ell^{\prime}}w_{\ell^{\prime}\ell
}\left[  f_{\ell}\left(  1-f_{\ell^{\prime}}\right)  -e^{\beta\left(
\epsilon_{\ell^{\prime}}-\epsilon_{\ell}\right)  }f_{\ell^{\prime}}\left(
1-f_{\ell}\right)  \right]  .\label{SBE-ep-original}%
\end{equation}

The argument about introducing the coordinate-shift into this equation is
similar to that in the case of static disorder, but is a little more involved
because $f_{\ell^{\prime}}$ appears in both the scattering-in and scattering-out terms. In the scattering-out term ($\ell
\rightarrow\ell^{\prime}$) of Eq. (\ref{SBE-ep-original}), the kinetic energy
of an electron in state $\ell^{\prime}$ after scattering out of state $\ell$
via absorbing (emitting) a phonon is $\epsilon_{\ell}\pm\hbar\omega_{q}%
+e\mathbf{E}\cdot\delta\mathbf{r}_{\ell^{\prime}\ell}$. In the scattering-in term
($\ell^{\prime}\rightarrow\ell$), the kinetic energy of an electron in state
$\ell^{\prime}$ before scattering into state $\ell$ via emitting (absorbing) a
phonon is $\epsilon_{\ell}%
\pm\hbar\omega_{q}-e\mathbf{E}\cdot\delta\mathbf{r}_{\ell\ell^{\prime}}$. Thus in
the linear response regime ($\epsilon_{\ell^{\prime}}=\epsilon_{\ell}\pm
\hbar\omega_{q}$), we have%
\begin{align}
&  \sum_{\ell^{\prime}}w_{\ell^{\prime}\ell}\left[  f_{\ell}\left(
1-f_{\ell^{\prime}}\right)  -e^{\beta\left(  \epsilon_{\ell^{\prime}}%
-\epsilon_{\ell}\right)  }f_{\ell^{\prime}}\left(  1-f_{\ell}\right)  \right]
\nonumber\\
%&  =\sum_{\ell^{\prime}}w_{\ell^{\prime}\ell}\left[  \left(  f_{\ell}%
%^{0}+g_{\ell}\right)  \left(  1-f_{\ell^{\prime}}^{0}-\delta f_{\ell^{\prime}%
%}\right)  -e^{\beta\left(  \epsilon_{\ell^{\prime}}-\epsilon_{\ell}\right)
%}\left(  f_{\ell^{\prime}}^{0}+\delta f_{\ell^{\prime}}\right)  \left(
%1-f_{\ell}^{0}-\delta f_{\ell}\right)  \right]  \nonumber\\
&  =\sum_{\ell^{\prime}}w_{\ell^{\prime}\ell}\left\{  \left(  f^{0}\left(
\epsilon_{\ell}\right)  +\delta f_{\ell}\right)  \left[  1-f^{0}\left(
\epsilon_{\ell^{^{\prime}}}+e\mathbf{E}\cdot\delta\mathbf{r}_{\ell^{\prime
}\ell}\right)  -\delta f_{\ell^{\prime}}\right]  \right.  \nonumber\\
&  \left.  -e^{\beta\left(  \epsilon_{\ell^{\prime}}-\epsilon_{\ell}\right)
}\left[  f^{0}\left(  \epsilon_{\ell^{\prime}}+e\mathbf{E}\cdot\delta
\mathbf{r}_{\ell^{\prime}\ell}\right)  +\delta f_{\ell^{\prime}}\right]
\left(  1-f_{\ell}^{0}-\delta f_{\ell}\right)  \right\}  \nonumber\\
&  =\sum_{\ell^{\prime}}w_{\ell^{\prime}\ell}\left\{  f^{0}\left(
\epsilon_{\ell}\right)  \left[  1-f^{0}\left(  \epsilon_{\ell^{\prime}%
}\right)  \right]  -e^{\beta\left(  \epsilon_{\ell^{\prime}}-\epsilon_{\ell
}\right)  }f^{0}\left(  \epsilon_{\ell^{\prime}}\right)  \left(
1-f^{0}\left(  \epsilon_{\ell}\right)  \right)  \right\}  \\
&  +\sum_{\ell^{\prime}}w_{\ell^{\prime}\ell}\left[  -f^{0}\left(
\epsilon_{\ell}\right)  -e^{\beta\left(  \epsilon_{\ell^{\prime}}%
-\epsilon_{\ell}\right)  }\left(  1-f^{0}\left(  \epsilon_{\ell}\right)
\right)  \right]  \frac{\partial f^{0}}{\partial\epsilon_{\ell^{\prime}}%
}e\mathbf{E}\cdot\delta\mathbf{r}_{\ell^{\prime}\ell}\nonumber\\
&  +\sum_{\ell^{\prime}}w_{\ell^{\prime}\ell}\left\{  \delta f_{\ell}\left[
1-f^{0}\left(  \epsilon_{\ell^{\prime}}\right)  \right]  -f^{0}\left(
\epsilon_{\ell}\right)  \delta f_{\ell^{\prime}}+e^{\beta\left(
\epsilon_{\ell^{\prime}}-\epsilon_{\ell}\right)  }\left[  f^{0}\left(
\epsilon_{\ell^{\prime}}\right)  \delta f_{\ell}-\delta f_{\ell^{\prime}%
}\left(  1-f^{0}\left(  \epsilon_{\ell}\right)  \right)  \right]  \right\}
+O\left(  \mathbf{E}^{2}\right),  \nonumber
\end{align}
where $\delta f$ is the out-of-equilibrium distribution.
On the right hand side of the last equality the first term is zero, and other
two terms can be simplified, leading to the following modified Bloch-Boltzmann equation
\begin{equation}
e\mathbf{E}\cdot\mathbf{v}_{\ell}^{0}\left(  -\frac{\partial f^{0}}%
{\partial\epsilon_{\ell}}\right)  =\sum_{\ell^{\prime}}w_{\ell^{\prime}\ell
}\left[  \delta f_{\ell}\frac{1-f^{0}\left(  \epsilon_{\ell^{\prime}}\right)
}{1-f^{0}\left(  \epsilon_{\ell}\right)  }-\delta f_{\ell^{\prime}}\frac
{f^{0}\left(  \epsilon_{\ell}\right)  }{f^{0}\left(  \epsilon_{\ell^{\prime}%
}\right)  }-\frac{f^{0}\left(  \epsilon_{\ell}\right)  }{f^{0}\left(
\epsilon_{\ell^{\prime}}\right)  }\frac{\partial f^{0}}{\partial\epsilon
_{\ell^{\prime}}}e\mathbf{E}\cdot\delta\mathbf{r}_{\ell^{\prime}\ell}\right]
.\label{SBE-ep}%
\end{equation}
By expressing $\delta f_{\ell}=g_{\ell}\left(  -\frac{\partial f^{0}}%
{\partial\epsilon_{\ell}}\right)  $, we arrive at Eq.~(\ref{SBE-new}) in the main text.

\section{Calculation details in the 2D massive Dirac model}

In the two-dimensional massive Dirac model, $\Omega_{k}=$ $-\frac{\Delta
v^{2}}{2(\Delta^{2}+\left(  vk\right)  ^{2})^{3/2}}$ is the Berry-curvature in
the positive band. Thus the side-jump velocity and the anomalous distribution
are given by%
\begin{equation}
v_{\bm k,y}^{\text{sj}}=-\frac{\Omega_{k}k_{x}}{\tau_{k}^{\text{sj}}}\text{
\ \ and \ \ \ }g_{\bm k}^{a}=-eE_{x}\Omega_{k}k_{y}\frac{\tau_{k}^{\text{tr}}%
}{\tau_{k}^{\text{sj}}}.
\end{equation}
By using the identity%
\begin{equation}
\frac{1-f^{0}\left(  \epsilon+\omega_{q}\right)  }{1-f^{0}\left(
\epsilon\right)  }N\left(  \omega_{q}\right)  +\frac{1-f^{0}\left(
\epsilon-\omega_{q}\right)  }{1-f^{0}\left(  \epsilon\right)  }\left[
N\left(  \omega_{q}\right)  +1\right]  =\frac{f^{0}\left(  \epsilon-\omega
_{q}\right)  -f^{0}\left(  \epsilon+\omega_{q}\right)  }{f^{0}\left(
\epsilon\right)  \left[  1-f^{0}\left(  \epsilon\right)  \right]  }N\left(
\omega_{q}\right)  \left[  N\left(  \omega_{q}\right)  +1\right]  ,
\end{equation}
the slight inelasticity of acoustic phonon scattering renders
\begin{equation}
\frac{1-f_{\bm k^{\prime}}^{0}}{1-f_{\bm k}^{0}}w_{\bm{k}^{\prime}%
\bm{k}}=\frac{2\pi}{\hbar}|\langle u_{\bm k^{\prime}}|u_{\bm k}\rangle
|^{2}\left\vert V_{\bm{k}^{\prime}\bm{k}}^{\text{o}}\right\vert ^{2}%
\frac{2\hbar\omega_{q}}{k_{B}T}N_{q}\left(  N_{q}+1\right)  \delta\left(
\epsilon_{k}-\epsilon_{k^{\prime}}\right)  ,
\end{equation}
where $q=2k\sin\frac{1}{2}\phi_{\bm{kk}^{\prime}}$. Thus%
\begin{equation}
\frac{\tau_{k}^{\text{tr}}}{\tau_{k}^{\text{sj}}}=\frac{\int d\phi
_{\bm{kk}^{\prime}}W_{\phi_{\bm{kk}^{\prime}}}\left(  1-\cos\phi
_{\bm{kk}^{\prime}}\right)  }{\int d\phi_{\bm{kk}^{\prime}}|\langle u_{\bm
k^{\prime}}|u_{\bm k}\rangle|^{2}W_{\phi_{\bm{kk}^{\prime}}}\left(  1-\cos
\phi_{\bm{kk}^{\prime}}\right)  },\label{ratio}%
\end{equation}
where%
\begin{equation}
W_{\phi_{\bm{kk}^{\prime}}}=\lambda^{2}k_{B}T\left(  \frac{\hbar\omega_{q}%
}{k_{B}T}\right)  ^{2}N_{q}\left(  N_{q}+1\right)  ,
\end{equation}
and $\lambda$ is the so-called electron-phonon coupling constant for the
deformation-potential treatment of the electron-phonon coupling \cite{Abrikosov,Kim2010}:
$2\left\vert V_{\bm{k}^{\prime
}\bm{k}}^{\text{o}}\right\vert ^{2}/\hbar\omega_{q}=\lambda^{2}$.

In the high-$T$ regime $W=\lambda^{2}k_{B}T$ is uniformly distributed on the Fermi circle,
and drops out of both the numerator
and denominator of $\tau_{k}^{\text{tr}}/\tau_{k}^{\text{sj}}$, thus
$\sigma_{\text{AH}}^{\text{sj}}$ takes the same $T$-independent value similar to that
due to scalar zero-range impurities. While at low temperatures the temperature
dependence of $N_{q}$ influences the integrals in $\tau_{k}^{\text{tr}%
}/\tau_{k}^{\text{sj}}$, and $\sigma_{\text{AH}}^{\text{sj}}$ becomes
$T$-dependent. In the low-$T$ limit $W/k_{B}T$ is highly peaked around
$\phi_{\bm{kk}^{\prime}}=0$ hence $|\langle u_{\bm k^{\prime}}|u_{\bm
k}\rangle|^{2}\rightarrow1$, $\tau_{k}^{\text{tr}}/\tau_{k}^{\text{sj}%
}\rightarrow1$ and $\sigma_{\text{AH}}^{\text{sj}}$ coincides with that due to
long-range scalar-impurities \cite{Xiao2007}.

\section{Generalized Bloch-Boltzmann formalism from the density matrix
approach}

To prove the validity of Eqs.~(\ref{SBE-new}) --~(\ref{sj velocity}) in the
main text, in the following two sections, we provide the microscopic
foundation for the Boltzmann formalism in weakly coupled electron-phonon
systems. Firstly, the density-matrix equation-of-motion approach
\cite{KL1957,Xiao2018KL} is applied to the many-particle density matrix for
the whole electron-phonon system \cite{Argyres1961}. The quantum Liouville
equation is analyzed in the occupation number representation perturbatively
with respect to the coupling parameter. Aside from the usual assumption that
the phonon system remains approximately in thermal equilibrium
\cite{Ziman1960,Holstein1964,Allen1978}, a basic statistical assumption is
needed, which is analogous to the assumption of molecular chaos made in
deriving the classical Boltzmann equation from the classical Liouville
equation \cite{Kardar}. We also show that the side jump contribution is
connected to the scattering-induced interband-coherence responses in the
microscopic transport theory, similar to the case of static disorder
\cite{Sinitsyn2008,Sinitsyn2006}. This clearly goes beyond the relaxation time
treatment where the effect of phonons is embodied only in an inelastic
lifetime of electrons \cite{Shitade2012}.

%As a bare perturbation theory, in the above density-matrix approach, a
%systematic renormalization procedure is absent. But it is expected naturally
%that the quantities appearing in the Boltzmann equation are physical
%(renormalized, e.g., RPA-type) ones \cite{Mahan}. This can only be verified by
%resorting to a renormalized many-body transport theory. Lyo \cite{Lyo1973}
%once extended Holstein's many-body electron-phonon transport formalism
%\cite{Holstein1964} to the anomalous Hall effect, ending up with a complicated
%theory having some gauge-dependent quantities as basic ingredients. We
%identify Lyo's equations corresponding to our generalized Bloch-Boltzmann
%formalism concerning coordinate-shift effects, laying a solid foundation for
%the latter.

For discussing problems in a quantum many-particle system, the second
quantized formalism is a common starting point. We introduce the notation
$\tilde{A}$ to denote the representation of an operator $\hat{A}$ in the
second-quantized formalism. For a single-particle operator, i.e., $\hat
{A}=\sum_{i}\hat{A}_{i}$ where $\hat{A}_{i}$ depends only on the dynamical
variables of the $i$-th carrier, we write $\tilde{A}=\sum_{\ell\ell^{\prime}%
}A_{\ell\ell^{\prime}}a_{\ell}^{\dag}a_{\ell^{\prime}}$ where $A_{\ell
\ell^{\prime}}$ is the corresponding matrix elements in the $\ell$
representation, $a_{\ell}^{\dag}$ and $a_{\ell}$ are the creation and
annihilation operators for the single-electron state $\left\vert
\ell\right\rangle $. The original version of Kohn-Luttinger density-matrix
approach \cite{KL1957} rests on the existence of a single-electron Hamiltonian
which contains all the information in the case of independent electrons
interacting with static disorder. In the case of dynamical disorder such as
phonons and magnons, as first pointed out by Argyres \cite{Argyres1961}, one
can apply the Kohn-Luttinger treatment to the many-body density matrix in the
occupation number representation for the whole system. Such a total
Hamiltonian reads%
\begin{equation}
\tilde{H}_{T}=\tilde{H}_{e}+\tilde{H}^{\prime}+\tilde{H}_{F}+\tilde{H}_{s},
\end{equation}
where $\tilde{H}_{e}=\sum_{mm^{\prime}}\left(  \hat{H}_{e}\right)
_{mm^{\prime}}a_{m}^{\dag}a_{m^{\prime}}$ is the electron Hamiltonian in the
absence of external electric fields and scattering, and $\tilde{H}_{F}%
=\sum_{mm^{\prime}}\left(  \hat{H}_{F}\right)  _{mm^{\prime}}a_{m}^{\dag
}a_{m^{\prime}}$ is the external-electric-field perturbation with $\hat{H}%
_{F}=\hat{H}_{1}e^{st}$ ($\hat{H}_{1}=-e\mathbf{E}\cdot\mathbf{\hat{r}}$)
turned on adiabatically from the remote past. The electric field is turned on
much more slowly than the scattering time ($s\rightarrow0^{+}$)
\cite{KL1957,Moore1967}. $\tilde{H}_{s}$ is the Hamiltonian of the scattering
system, and $\hat{H}^{\prime}=\lambda\hat{V}$ is the interaction of electrons
with the scattering system, where $\lambda$\ is a dimensionless parameter used
for analyzing the order in the perturbative analysis and is set to 1
eventually. $\left(  \hat{H}^{\prime}\right)  _{mm^{\prime}}$ is still an
operator in the Hilbert space of the scattering system. In the occupation
number representation $\left\{  \left\vert nN\right\rangle \right\}  $,
$\tilde{H}_{e}\left\vert nN\right\rangle =\sum_{\ell}\epsilon_{\ell}n_{\ell
}\left\vert nN\right\rangle =E_{n}\left\vert nN\right\rangle $ and $\tilde
{H}_{s}\left\vert nN\right\rangle =E_{N}\left\vert nN\right\rangle $.
Hereafter we set $E_{nN}\equiv E_{n}+E_{N}$, $n$ and $N$ are the many-particle
state indices for the electron system and scattering system, respectively.
$\hat{n}_{\ell}=a_{\ell}^{\dag}a_{\ell}$, and its eigenvalue $n_{\ell}$
denotes the electron number on the Bloch state marked by the index $\ell$ with
single-electron eigenenergy $\epsilon_{\ell}$. In the linear response regime
the total many-particle density matrix reads%
\begin{equation}
\tilde{\rho}_{T}=\tilde{\rho}+\tilde{F}e^{st},
\end{equation}
where $\tilde{\rho}$ is the equilibrium many-particle density matrix for the
whole system, and $\tilde{F}$ is linear in the electric field. The quantum
Liouville equation
\begin{equation}
i\hbar\frac{\partial}{\partial t}\tilde{\rho}_{T}=\left[  \tilde{H}_{T}%
,\tilde{\rho}_{T}\right]
\end{equation}
becomes $i\hbar s\tilde{F}=\left[  \tilde{H}_{0}+\tilde{H}_{s}+\tilde
{H}^{\prime},\tilde{F}\right]  +\left[  \tilde{H}_{1},\tilde{\rho}\right] . $
In the occupation number representation $\left\{  \left\vert nN\right\rangle
\right\}  $ one has
\begin{equation}
\left(  E_{nN}-E_{n^{\prime}N^{\prime}}-i\hbar s\right)  \tilde{F}%
_{nN,n^{\prime}N^{\prime}}=\sum_{n^{\prime\prime}N^{\prime\prime}}\left(
\tilde{F}_{nN,n^{\prime\prime}N^{\prime\prime}}\tilde{H}_{n^{\prime\prime
}N^{\prime\prime},n^{\prime}N^{\prime}}^{^{\prime}}-\tilde{H}_{nN,n^{\prime
\prime}N^{\prime\prime}}^{^{\prime}}\tilde{F}_{n^{\prime\prime}N^{\prime
\prime},n^{\prime}N^{\prime}}\right)  +\tilde{C}_{nN,n^{\prime}N^{\prime}},
\label{QLE}%
\end{equation}
where $\tilde{C}_{nN,n^{\prime}N^{\prime}}\equiv\left[  \tilde{\rho},\tilde
{H}_{1}\right]  _{nN,n^{\prime}N^{\prime}}$. Hereafter we sometimes use the
notation $L=nN$, $L^{^{\prime}}=n^{^{\prime}}N^{^{\prime}}$ to simplify expressions.

The linear response of an observable $A$ is $\delta A=\text{Tr}\left(
\tilde{F}\tilde{A}\right)  =\sum_{LL^{^{\prime}}}\tilde{F}_{LL^{^{\prime}}%
}\tilde{A}_{L^{^{\prime}}L}=\sum_{L}\tilde{F}_{L}\tilde{A}_{LL}+\sum
_{LL^{^{\prime}}}^{^{\prime}}\tilde{F}_{LL^{^{\prime}}}\tilde{A}_{L^{^{\prime
}}L},$ where Tr denotes the trace operation in the occupation-number space,
and the notation $\sum^{^{\prime}}$ means that all the index equalities in the
summation are avoided. Here we first outline the main results of the following
detailed derivation. The linear response of the velocity of electrons is
\begin{equation}
\delta\mathbf{v}=\text{Tr}\left(  \tilde{F}\mathbf{\tilde{v}}\right)
=\sum_{L}\tilde{F}_{L}\mathbf{\tilde{v}}_{LL}+\sum_{LL^{\prime}}^{\prime
}\tilde{F}_{LL^{\prime}}\mathbf{\tilde{v}}_{L^{\prime}L}.
\end{equation}
To obtain $\tilde{F}_{L}$ and $\tilde{F}_{LL^{^{\prime}}}$ in the weakly
coupled system we make a perturbative analysis of Eq. (\ref{QLE}) with respect
to the coupling parameter. The off-diagonal elements $\tilde{F}_{LL^{\prime}}$
can be expressed in terms of the diagonal ones $\tilde{F}_{L}$, resulting in
an equation for $\tilde{F}_{L}$. Because by definition $f_{\ell}^{0}=$
Tr$\left(  \hat{n}_{\ell}\tilde{\rho}\right)  =\sum_{L}n_{\ell}\tilde{\rho
}_{L}$ and
\begin{equation}
\delta f_{\ell}=\text{Tr}\left(  \hat{n}_{\ell}\tilde{F}\right)  =\sum
_{L}n_{\ell}\tilde{F}_{L},\label{link-1}%
\end{equation}
we derive the modified Bloch-Boltzmann equation (10) of the main text based on
the equation for $\tilde{F}_{L}$. According to Eq. (\ref{link-1}) one has
\begin{equation}
\sum_{L}\tilde{F}_{L}\mathbf{\tilde{v}}_{LL}=\sum_{L}\tilde{F}_{L}\sum_{\ell
}\mathbf{v}_{\ell}^{0}n_{\ell}=\sum_{\ell}\delta f_{\ell}\mathbf{v}_{\ell}%
^{0}.
\end{equation}
Whereas $\sum_{LL^{\prime}}^{\prime}\tilde{F}_{LL^{\prime}}\mathbf{\tilde{v}%
}_{L^{\prime}L}$ is proven to yield the transport contributions from the
Berry-curvature anomalous velocity and the side-jump velocity:%
\begin{equation}
\sum_{LL^{\prime}}\ ^{\prime}\tilde{F}_{LL^{\prime}}\mathbf{\tilde{v}%
}_{L^{\prime}L}=\sum_{\ell}f_{\ell}^{0}\mathbf{v}_{\ell}^{\text{bc}}%
+\sum_{\ell}\delta f_{\ell}^{n}\left[  \sum_{\ell^{\prime}}\frac
{1-f_{\ell^{\prime}}^{0}}{1-f_{\ell}^{0}}w_{\ell^{\prime}\ell}\delta
\mathbf{r}_{\ell^{\prime}\ell}\right]  ,
\end{equation}
where $\delta\mathbf{r}_{\ell^{\prime}\ell}$ is given by Eq. (3) of the main
text. We also show that the side-jump velocity $\mathbf{v}_{\ell}^{\text{sj}%
}=\sum_{\ell^{\prime}}\frac{1-f_{\ell^{\prime}}^{0}}{1-f_{\ell}^{0}}%
w_{\ell^{\prime}\ell}\delta\mathbf{r}_{\ell^{\prime}\ell}$ arises from
scattering-induced interband-coherence, so does the anomalous distribution
function $g_{\ell}^{a}$ (Eqs.~(\ref{SBE-a}) and (\ref{sj velocity})).

\subsection{Perturbative analysis of the quantum Liouville equation}

We split the quantum Liouville equation into diagonal and off-diagonal parts
in the $\left\vert nN\right\rangle $-representation:%
\begin{align}
\left(  E_{nN}+\tilde{H}_{nN}^{^{\prime}}-E_{n^{\prime}N^{\prime}}-\tilde
{H}_{n^{\prime}N^{\prime}}^{^{\prime}}-i\hbar s\right)  \tilde{F}%
_{nN,n^{\prime}N^{\prime}} &  =\sum_{n^{\prime\prime}N^{\prime\prime}}%
^{\prime}\left(  \tilde{F}_{nN,n^{\prime\prime}N^{\prime\prime}}\tilde
{H}_{n^{\prime\prime}N^{\prime\prime},n^{\prime}N^{\prime}}^{^{\prime}}%
-\tilde{H}_{nN,n^{\prime\prime}N^{\prime\prime}}^{^{\prime}}\tilde
{F}_{n^{\prime\prime}N^{\prime\prime},n^{\prime}N^{\prime}}\right)
\nonumber\\
&  +\left(  \tilde{F}_{nN}-\tilde{F}_{n^{\prime}N^{\prime}}\right)  \tilde
{H}_{nN,n^{\prime}N^{\prime}}^{^{\prime}}+\tilde{C}_{nN,n^{\prime}N^{\prime}%
},\label{KL-offdiagonal}%
\end{align}
for $nN\neq n^{^{\prime}}N^{^{\prime}}$, and%
\begin{equation}
-i\hbar s\tilde{F}_{nN}=\sum_{n^{\prime}N^{\prime}}^{\prime}\left(  \tilde
{F}_{nN,n^{\prime}N^{\prime}}\tilde{H}_{n^{\prime}N^{\prime},nN}^{^{\prime}%
}-\tilde{H}_{nN,n^{\prime}N^{\prime}}^{^{\prime}}\tilde{F}_{n^{\prime
}N^{\prime},nN}\right)  +\tilde{C}_{nN}.\label{KL-diagonal}%
\end{equation}
According to the spirit of the Boltzmann theory, the first-order energy shift
$\tilde{H}_{nN}^{^{\prime}}$ is incorporated into the renormalization of the
band energy and henceforth neglected \cite{KL1957,Xiao2018KL}. To solve these
two equations in the weak coupling regime we make the standard order-by-order
analysis with respect to the coupling parameter of the interaction with
disorder:%
\begin{align}
\tilde{F}_{nN} &  =\tilde{F}_{nN}^{\left(  -2\right)  }+\tilde{F}%
_{nN}^{\left(  -1\right)  }+\tilde{F}_{nN}^{\left(  0\right)  }%
+...,\nonumber\\
\tilde{F}_{nN,n^{\prime}N^{\prime}} &  =\tilde{F}_{nN,n^{\prime}N^{\prime}%
}^{\left(  -1\right)  }+\tilde{F}_{nN,n^{\prime}N^{\prime}}^{\left(  0\right)
}+\tilde{F}_{nN,n^{\prime}N^{\prime}}^{\left(  1\right)  }...,\\
\tilde{C}_{nN,n^{\prime}N^{\prime}} &  =\tilde{C}_{nN,n^{\prime}N^{\prime}%
}^{\left(  0\right)  }+\tilde{C}_{nN,n^{\prime}N^{\prime}}^{\left(  1\right)
}+\tilde{C}_{nN,n^{\prime}N^{\prime}}^{\left(  2\right)  }+...\nonumber
\end{align}
Hereafter the superscript $\left(  i\right)  $ denotes the order in $\lambda$.

For Eq. (\ref{KL-offdiagonal}) one can obtain: in $O\left(  \lambda
^{-1}\right)  $%
\begin{equation}
\left(  E_{nN}-E_{n^{\prime}N^{\prime}}-i\hbar s\right)  \tilde{F}%
_{nN,n^{\prime}N^{\prime}}^{\left(  -1\right)  }=\left[  \tilde{F}%
_{nN}^{\left(  -2\right)  }-\tilde{F}_{n^{\prime}N^{\prime}}^{\left(
-2\right)  }\right]  \tilde{H}_{nN,n^{\prime}N^{\prime}}^{^{\prime}},
\end{equation}
in $O\left(  \lambda^{0}\right)  $%
\begin{align}
\left[  E_{nN}-E_{n^{\prime}N^{\prime}}-i\hbar s\right]  \tilde{F}%
_{nN,n^{\prime}N^{\prime}}^{\left(  0\right)  } &  =\sum_{n^{\prime\prime
}N^{\prime\prime}}^{\prime}\left[  \tilde{F}_{nN,n^{\prime\prime}%
N^{\prime\prime}}^{\left(  -1\right)  }\tilde{H}_{n^{\prime\prime}%
N^{\prime\prime},n^{\prime}N^{\prime}}^{^{\prime}}-\tilde{H}_{nN,n^{\prime
\prime}N^{\prime\prime}}^{^{\prime}}\tilde{F}_{n^{\prime\prime}N^{\prime
\prime},n^{\prime}N^{\prime}}^{\left(  -1\right)  }\right]  \nonumber\\
&  +\left[  \tilde{F}_{nN}^{\left(  -1\right)  }-\tilde{F}_{n^{\prime
}N^{\prime}}^{\left(  -1\right)  }\right]  \tilde{H}_{nN,n^{\prime}N^{\prime}%
}^{^{\prime}}+\tilde{C}_{nN,n^{\prime}N^{\prime}}^{\left(  0\right)  },
\end{align}
in $O\left(  \lambda\right)  $%
\begin{align}
\left[  E_{nN}-E_{n^{\prime}N^{\prime}}-i\hbar s\right]  \tilde{F}%
_{nN,n^{\prime}N^{\prime}}^{\left(  1\right)  } &  =\sum_{n^{\prime\prime
}N^{\prime\prime}}^{\prime}\left[  \tilde{F}_{nN,n^{\prime\prime}%
N^{\prime\prime}}^{\left(  0\right)  }\tilde{H}_{n^{\prime\prime}%
N^{\prime\prime},n^{\prime}N^{\prime}}^{^{\prime}}-\tilde{H}_{nN,n^{\prime
\prime}N^{\prime\prime}}^{^{\prime}}\tilde{F}_{n^{\prime\prime}N^{\prime
\prime},n^{\prime}N^{\prime}}^{\left(  0\right)  }\right]  \nonumber\\
&  +\left[  \tilde{F}_{nN}^{\left(  0\right)  }-\tilde{F}_{n^{\prime}%
N^{\prime}}^{\left(  0\right)  }\right]  \tilde{H}_{nN,n^{\prime}N^{\prime}%
}^{^{\prime}}+\tilde{C}_{nN,n^{\prime}N^{\prime}}^{\left(  1\right)  }.
\end{align}
For Eq. (\ref{KL-offdiagonal}) one can obtain: in $O\left(  \lambda
^{0}\right)  $%
\begin{equation}
0=\sum_{n^{\prime}N^{\prime}}^{\prime}\left[  \tilde{F}_{nN,n^{\prime
}N^{\prime}}^{\left(  -1\right)  }\tilde{H}_{n^{\prime}N^{\prime}%
,nN}^{^{\prime}}-\tilde{H}_{nN,n^{\prime}N^{\prime}}^{^{\prime}}\tilde
{F}_{n^{\prime}N^{\prime},nN}^{\left(  -1\right)  }\right]  +\tilde{C}%
_{nN}^{\left(  0\right)  },
\end{equation}
in $O\left(  \lambda\right)  $%
\begin{equation}
0=\sum_{n^{\prime}N^{\prime}}^{\prime}\left[  \tilde{F}_{nN,n^{\prime
}N^{\prime}}^{\left(  0\right)  }\tilde{H}_{n^{\prime}N^{\prime},nN}%
^{^{\prime}}-\tilde{H}_{nN,n^{\prime}N^{\prime}}^{^{\prime}}\tilde
{F}_{n^{\prime}N^{\prime},nN}^{\left(  0\right)  }\right]  +\tilde{C}%
_{nN}^{\left(  1\right)  },
\end{equation}
in $O\left(  \lambda^{2}\right)  $%
\begin{equation}
0=\sum_{n^{\prime}N^{\prime}}^{\prime}\left[  \tilde{F}_{nN,n^{\prime
}N^{\prime}}^{\left(  1\right)  }\tilde{H}_{n^{\prime}N^{\prime},nN}%
^{^{\prime}}-\tilde{H}_{nN,n^{\prime}N^{\prime}}^{^{\prime}}\tilde
{F}_{n^{\prime}N^{\prime},nN}^{\left(  1\right)  }\right]  +\tilde{C}%
_{nN}^{\left(  2\right)  }.
\end{equation}

For simplicity we assume the bosonic quasi-particles of the dynamical
scattering systems, e.g., phonons and/or magnons, can be approximately thought
to be in thermal equilibrium. Although this standard assumption after F. Bloch
\cite{Ziman1960} can only be clearly justified at high temperatures, it was
shown to work well in many cases beyond that regime
\cite{Ziman1960,Sarma1992,Kim2010}. Here we adopt it to simplify the
derivation (which is still quite tedious even after making this assumption).

The off-diagonal (with respect to $L$) elements $\tilde{F}_{LL^{\prime}}$ can
be expressed in terms of the diagonal ones $\tilde{F}_{L}$, and $\tilde{F}%
_{L}$ are related to the diagonal (in the single-electron Bloch
representation) elements of the single-electron density matrix (Eq.
(\ref{link-1})). Thus the Bloch-Boltzmann theory formulated in the
single-electron Bloch representation can be derived from the microscopic
transport theory presented in the occupation number representation.

\subsection{Perturbative calculation of $C_{LL^{^{\prime}}}$}

Applying the Karplus-Schwinger expansion \cite{Karplus1948}%
\begin{equation}
e^{\tilde{A}+\tilde{B}}=e^{\tilde{A}}+\int_{0}^{1}d\lambda e^{\left(
1-\lambda\right)  \tilde{A}}\tilde{B}e^{\lambda\tilde{A}}+\int_{0}^{1}d\lambda
e^{\left(  1-\lambda\right)  \tilde{A}}\tilde{B}e^{\lambda\tilde{A}}\int
_{0}^{\lambda}d\lambda^{^{\prime}}e^{-\lambda^{^{\prime}}\tilde{A}}\tilde
{B}e^{\lambda^{^{\prime}}\tilde{A}}+...
\end{equation}
up to the second order of $B$ one can calculate the equilibrium density matrix
$\tilde{\rho}=Z^{-1}e^{\tilde{A}+\tilde{B}}$ ($\tilde{A}=-\beta\left(
\tilde{H}_{e}-\mu\tilde{N}_{e}+\tilde{H}_{s}\right)  $, $\tilde{B}%
=-\beta\tilde{H}^{^{\prime}}$) in weakly coupled systems. The partition
function is given by $Z^{-1}\simeq Z_{0}^{-1}\left(  1+\gamma\right)  $, where
$Z_{0}=\sum_{L}e^{A_{L}}$ and $\gamma\sim o\left(  B^{2}\right)  $. We have
($\tilde{\rho}^{\left(  0\right)  }=Z_{0}^{-1}e^{\tilde{A}}$)%
\begin{align*}
\tilde{C}^{\left(  0\right)  } &  \equiv\left[  \tilde{\rho}^{\left(
0\right)  },\tilde{H}_{1}\right]  =Z_{0}^{-1}\left(  -e\mathbf{E}\right)
\cdot\exp\left(  -\beta\tilde{H}_{s}\right)  \left[  \exp\left(  -\beta
\sum_{j}\hat{H}_{e}\left(  j\right)  \right)  ,\sum_{i}\mathbf{\hat{r}}%
_{i}\right]  \\
&  =\left(  -e\mathbf{E}\right)  \cdot\tilde{\rho}^{\left(  0\right)  }%
\sum_{\ell\ell^{^{\prime}}}\exp\left(  \beta\epsilon_{\ell}\right)  \left[
\exp\left(  -\beta\hat{H}_{e}\right)  ,\mathbf{\hat{r}}\right]  _{\ell
\ell^{^{\prime}}}a_{\ell}^{\dag}a_{\ell^{^{\prime}}}\\
&  =i\tilde{\rho}^{\left(  0\right)  }e\mathbf{E}\cdot\left\{  \sum_{\ell
\ell^{^{\prime}}}^{\prime}\mathbf{J}_{\ell\ell^{^{\prime}}}\left[  \exp\left(
-\beta\left(  \epsilon_{\ell^{^{\prime}}}-\epsilon_{\ell}\right)  \right)
-1\right]  a_{\ell}^{\dag}a_{\ell^{^{\prime}}}+\left(  -\beta\right)
\sum_{\ell}\frac{\partial\epsilon_{\ell}}{\partial\mathbf{k}}\hat{n}_{\ell
}\right\}  ,
\end{align*}
then%
\begin{equation}
\tilde{C}_{nN,n^{\prime}N^{\prime}}^{\left(  0\right)  }=ie\mathbf{E}%
\cdot\left[  \sum_{\ell\ell^{^{\prime}}}^{\prime}\mathbf{J}_{\ell
\ell^{^{\prime}}}\left(  e^{-\beta\left(  \epsilon_{\ell^{^{\prime}}}%
-\epsilon_{\ell}\right)  }-1\right)  \tilde{\rho}_{nN}^{\left(  0\right)
}\left(  a_{\ell}^{\dag}a_{\ell^{^{\prime}}}\right)  _{n,n^{\prime}}\left(
1-\delta_{n,n^{\prime}}\right)  +\left(  -\beta\right)  \sum_{\ell}%
\frac{\partial\epsilon_{\ell}}{\partial\mathbf{k}}n_{\ell}\tilde{\rho}%
_{nN}^{\left(  0\right)  }\delta_{n,n^{\prime}}\right]  \delta_{N,N^{\prime}%
}.\label{C-0}%
\end{equation}
Next we look at%
\begin{align*}
\tilde{C}^{\left(  1\right)  } &  \equiv\left[  \tilde{\rho}^{\left(
1\right)  },\tilde{H}_{1}\right]  =\frac{1}{Z_{0}}\left[  \int_{0}^{1}d\lambda
e^{\left(  1-\lambda\right)  \tilde{A}}\tilde{B}e^{\lambda\tilde{A}},\tilde
{H}_{1}\right]  \\
&  =\frac{1}{Z_{0}}\int_{0}^{1}d\lambda\sum_{\ell\ell^{^{\prime}}}\left\{
\begin{array}
[c]{c}%
e^{\left(  1-\lambda\right)  \left(  \tilde{H}_{e}+\tilde{H}_{s}\right)
}\tilde{H}^{^{\prime}}e^{\lambda\left(  \tilde{H}_{e}+\tilde{H}_{s}\right)
}e^{-\lambda A_{\ell}}\left[  e^{\lambda\hat{H}_{e}},\hat{H}_{1}\right]
_{\ell\ell^{^{\prime}}}a_{\ell}^{\dag}a_{\ell^{^{\prime}}}\\
+e^{\left(  1-\lambda\right)  \left(  \tilde{H}_{e}+\tilde{H}_{s}\right)
}\left[  \hat{H}^{^{\prime}},\hat{H}_{1}\right]  _{\ell\ell^{^{\prime}}%
}a_{\ell}^{\dag}a_{\ell^{^{\prime}}}e^{\lambda\left(  \tilde{H}_{e}+\tilde
{H}_{s}\right)  }\\
+e^{\left(  1-\lambda\right)  \left(  \tilde{H}_{e}+\tilde{H}_{s}\right)
}e^{-\left(  1-\lambda\right)  A_{\ell}}\left[  e^{\left(  1-\lambda\right)
\hat{H}_{e}},\hat{H}_{1}\right]  _{\ell\ell^{^{\prime}}}a_{\ell}^{\dag}%
a\tilde{H}^{^{\prime}}e^{\lambda\left(  \tilde{H}_{e}+\tilde{H}_{s}\right)  }%
\end{array}
\right\}  .
\end{align*}
There are so many terms that one should have some guiding principle to
simplify the analysis. According to the insight we obtained in the discussion
of static-disorder case \cite{Xiao2018KL}, some trivial renormalization
effects can be neglected and only the diagonal (in the Bloch representation
for electrons) elements of electric-field perturbation survive in the final
contribution to $\tilde{C}_{L}^{^{\prime\prime}}$, which appears in the
following Eq. (\ref{F0}) as an anomalous driving term \cite{KL1957,Xiao2018KL}%
. Thus, we obtain
\begin{equation}
\tilde{C}_{nN,n^{\prime}N^{\prime}}^{\left(  1\right)  }=\sum_{\ell
\ell^{^{\prime}}}^{\prime}ie\mathbf{E\cdot}\left[  \left(  \mathbf{J}_{\ell
}-\mathbf{J}_{\ell^{^{\prime}}}\right)  H_{\ell N,\ell^{^{\prime}}N^{\prime}%
}^{^{\prime}}+iH_{\ell N,\ell^{^{\prime}}N^{\prime}}^{^{\prime}}%
\mathbf{\hat{D}}\arg H_{\ell N,\ell^{^{\prime}}N^{\prime}}^{^{\prime}}\right]
\left(  a_{\ell}^{\dag}a_{\ell^{^{\prime}}}\right)  _{n,n^{\prime}}%
\frac{\tilde{\rho}_{n^{\prime}N^{\prime}}^{\left(  0\right)  }-\tilde{\rho
}_{nN}^{\left(  0\right)  }}{E_{n^{\prime}N^{\prime}}-E_{nN}},\label{C-1}%
\end{equation}
where $\mathbf{\hat{D}}=\partial_{\mathbf{k}}+\partial_{\mathbf{k}^{\prime}}$,
$\mathbf{J}_{\ell}=\langle u_{\ell}|\partial_{\mathbf{k}}|u_{\ell}\rangle$ and
$\mathbf{J}_{\ell\ell^{\prime}}=\delta_{\mathbf{kk}^{\prime}}\langle u_{\ell
}|\partial_{\mathbf{k}}|u_{\ell^{\prime}}\rangle$. Meanwhile the anomalous
driving term that will appear in Eq. (\ref{F0})
\begin{equation}
\tilde{C}_{nN}^{^{^{\prime\prime}}}=\sum_{n^{^{\prime}}N^{^{\prime}}}^{\prime
}\left[  \frac{\tilde{C}_{nN,n^{^{\prime}}N^{^{\prime}}}^{\left(  1\right)
}\tilde{H}_{n^{^{\prime}}N^{^{\prime}},nN}^{^{^{\prime}}}}{d_{nN,n^{^{\prime}%
}N^{^{\prime}}}^{-}}-c.c.\right]  \label{anomalous driving}%
\end{equation}
only contains nontrivial correction to the driving term of the transport
equation with $\tilde{C}_{LL^{\prime}}^{\left(  1\right)  }$ given by Eq.
(\ref{C-1}). One can verify that $\left(  \tilde{C}_{nN,n^{\prime}N^{\prime}%
}^{\left(  1\right)  }\right)  ^{\ast}=-\tilde{C}_{n^{\prime}N^{\prime}%
,nN}^{\left(  1\right)  }$. Henceforth $d_{nN,n^{\prime}N^{\prime}}^{\pm
}\equiv E_{nN}-E_{n^{\prime}N^{\prime}}\pm i\hbar s$. In the above derivation
we used $\left[  \hat{\rho},\mathbf{r}\right]  _{\ell\ell^{^{\prime}}}%
=-i\sum_{\ell^{^{\prime\prime}}}\left(  \mathbf{J}_{\ell\ell^{^{\prime\prime}%
}}\rho_{\ell^{^{\prime\prime}}\ell^{^{\prime}}}-\rho_{\ell\ell^{^{\prime
\prime}}}\mathbf{J}_{\ell^{^{\prime\prime}}\ell^{^{\prime}}}\right)
-i\mathbf{\hat{D}}\rho_{\ell\ell^{^{\prime}}}$ for $\ell\neq\ell^{^{\prime}}$
and $\left[  \hat{\rho},\mathbf{r}\right]  _{\ell\ell}=-i\sum_{\ell^{^{\prime
}}}\left(  \mathbf{J}_{\ell\ell^{^{\prime}}}\rho_{\ell^{^{\prime}}\ell}%
-\rho_{\ell\ell^{^{\prime}}}\mathbf{J}_{\ell^{^{\prime}}\ell}\right)
-i\frac{\partial}{\partial\mathbf{k}}\rho_{\ell\ell}$.

\subsection{Conventional Bloch-Boltzmann equation}

In the zeroth order of electron-disorder interaction one has
\begin{equation}
0=\tilde{C}_{L}^{\left(  0\right)  }+i\hbar\sum_{L^{^{\prime}}}\tilde{\omega
}_{LL^{^{\prime}}}^{\left(  2\right)  }\left[  \tilde{F}_{L}^{\left(
-2\right)  }-\tilde{F}_{L^{^{\prime}}}^{\left(  -2\right)  }\right]
\end{equation}
with $\tilde{\omega}_{LL^{^{\prime}}}^{\left(  2\right)  }=\frac{2\pi}{\hbar
}\left\vert \tilde{H}_{LL^{^{\prime}}}^{^{\prime}}\right\vert ^{2}%
\delta\left(  E_{nN}-E_{n^{\prime}N^{\prime}}\right)  $. Then
\begin{equation}
0=\sum_{nN}n_{\ell}\tilde{C}_{nN}^{\left(  0\right)  }+2\pi i\sum
_{nN,n^{\prime}N^{\prime}}^{\prime}\left\vert \tilde{H}_{nN,n^{\prime
}N^{\prime}}^{^{\prime}}\right\vert ^{2}\delta\left(  E_{nN}-E_{n^{\prime
}N^{\prime}}\right)  \left(  n_{\ell}-n_{\ell}^{\prime}\right)  \tilde{F}%
_{nN}^{\left(  -2\right)  },
\end{equation}
where%
\begin{align}
\sum_{nN}n_{\ell}\tilde{C}_{nN}^{\left(  0\right)  } &  =ie\mathbf{E}%
\cdot\left(  -\beta\right)  \sum_{\ell^{^{\prime}}}\frac{\partial
\epsilon_{\ell^{^{\prime}}}}{\partial\mathbf{k}^{^{\prime}}}\sum_{nN}n_{\ell
}n_{\ell^{^{\prime}}}\tilde{\rho}_{nN}^{\left(  0\right)  }=\left(
-\beta\right)  ie\mathbf{E}\cdot\sum_{\ell^{^{\prime}}}\frac{\partial
\epsilon_{\ell^{^{\prime}}}}{\partial\mathbf{k}^{^{\prime}}}\sum_{n}n_{\ell
}n_{\ell^{^{\prime}}}\tilde{\rho}_{n}^{\left(  0\right)  }\nonumber\\
&  =ie\mathbf{E}\cdot\frac{\partial\epsilon_{\ell}}{\partial\mathbf{k}}\left(
-\beta\right)  f_{\ell}^{0}\left(  1-f_{\ell}^{0}\right)  =ie\mathbf{E}%
\cdot\frac{\partial\epsilon_{\ell}}{\partial\mathbf{k}}\frac{\partial f_{\ell
}^{0}}{\partial\epsilon_{\ell}}=ie\mathbf{E}\cdot\frac{\partial f_{\ell}^{0}%
}{\partial\mathbf{k}},
\end{align}
and
\begin{align*}
&  2\pi i\sum_{nN,n^{\prime}N^{\prime}}^{\prime}\left\vert \tilde
{H}_{nN,n^{\prime}N^{\prime}}^{^{\prime}}\right\vert ^{2}\delta\left(
E_{nN}-E_{n^{\prime}N^{\prime}}\right)  \left(  n_{k}-n_{k}^{\prime}\right)
\tilde{F}_{nN}^{\left(  -2\right)  }\\
&  =2\pi i\sum_{nN,n^{\prime}N^{\prime}}\sum_{\ell\ell^{^{\prime}}}^{\prime
}\left\vert H_{\ell N,\ell^{^{\prime}}N^{\prime}}^{^{\prime}}\right\vert
^{2}n_{\ell}\left(  1-n_{\ell^{^{\prime}}}\right)  \delta_{n_{\ell}-1=n_{\ell
}^{^{\prime}}}\delta_{n_{\ell^{^{\prime}}}+1=n_{\ell^{^{\prime}}}^{^{\prime}}%
}\delta\left(  E_{N}-E_{N^{\prime}}+\epsilon_{\ell}-\epsilon_{\ell^{^{\prime}%
}}\right)  \left(  n_{k}-n_{k}^{\prime}\right)  \tilde{F}_{nN}^{\left(
-2\right)  }\\
&  =i\hbar\sum_{nN,N^{\prime}}\sum_{\ell^{^{\prime}}}^{\prime}\left[
\omega_{kN,\ell^{^{\prime}}N^{\prime}}^{2s}n_{k}\left(  1-n_{\ell^{^{\prime}}%
}\right)  -\omega_{\ell^{^{\prime}}N,kN^{\prime}}^{2s}n_{\ell^{^{\prime}}%
}\left(  1-n_{k}\right)  \right]  \tilde{F}_{nN}^{\left(  -2\right)  }.
\end{align*}
In the derivation one uses%
\begin{equation}
\left(  a_{\ell}^{\dag}a_{\ell^{^{\prime}}}\right)  _{n,n^{\prime}}\left(
a_{k^{\prime}}^{\dag}a_{k}\right)  _{n^{\prime},n}=\delta_{k\ell}%
\delta_{k^{\prime}\ell^{^{\prime}}}n_{\ell}\left(  1-n_{\ell^{^{\prime}}%
}\right)  \delta_{n_{\ell}-1=n_{\ell}^{^{\prime}}}\delta_{n_{\ell^{^{\prime}}%
}+1=n_{\ell^{^{\prime}}}^{^{\prime}}}.\label{operator-2}%
\end{equation}
Thus we obtain \cite{Argyres1961}
\begin{equation}
e\mathbf{E}\cdot\frac{\partial f_{\ell}^{0}}{\hbar\partial\mathbf{k}}%
+\sum_{nN,N^{\prime}}\sum_{\ell^{^{\prime}}}^{\prime}\left[  \omega_{\ell
N,\ell^{^{\prime}}N^{\prime}}^{2s}n_{\ell}\left(  1-n_{\ell^{^{\prime}}%
}\right)  -\omega_{\ell^{^{\prime}}N,\ell N^{\prime}}^{2s}n_{\ell^{^{\prime}}%
}\left(  1-n_{\ell}\right)  \right]  \tilde{F}_{nN}^{\left(  -2\right)  }=0,
\end{equation}
where $\omega_{\ell N,\ell^{^{\prime}}N^{\prime}}^{2s}=\frac{2\pi}{\hbar
}\left\vert H_{\ell N,\ell^{^{\prime}}N^{\prime}}^{^{\prime}}\right\vert
^{2}\delta\left(  E_{N}-E_{N^{\prime}}+\epsilon_{\ell}-\epsilon_{\ell
^{^{\prime}}}\right)  $. Since the bosonic quasi-particles of the dynamical
scattering systems (e.g., phonons or magnons) are assumed to remain in
equilibrium, we introduce the following assumption for factorizing the entire
many-particle density matrix \cite{Argyres1961}:
\begin{equation}
\tilde{F}_{nN}^{\left(  -2\right)  }=P_{N}^{\left(  0\right)  }\tilde{F}%
_{n}^{\left(  -2\right)  },\label{statistical-1}%
\end{equation}
then%
\[
\sum_{nN,N^{\prime}}\sum_{\ell^{^{\prime}}}^{\prime}\left[  \omega_{\ell
N,\ell^{^{\prime}}N^{\prime}}^{2s}n_{\ell}\left(  1-n_{\ell^{^{\prime}}%
}\right)  -\omega_{\ell^{^{\prime}}N,\ell N^{\prime}}^{2s}n_{\ell^{^{\prime}}%
}\left(  1-n_{\ell}\right)  \right]  \tilde{F}_{nN}^{\left(  -2\right)  }%
=\sum_{\ell^{^{\prime}}}^{\prime}\sum_{n}\left[  \omega_{\ell^{^{\prime}}\ell
}^{\left(  2\right)  }n_{\ell}\left(  1-n_{\ell^{^{\prime}}}\right)
-\omega_{\ell\ell^{^{\prime}}}^{\left(  2\right)  }n_{\ell^{^{\prime}}}\left(
1-n_{\ell}\right)  \right]  \tilde{F}_{n}^{\left(  -2\right)  },
\]
where%
\begin{align}
\omega_{\ell^{^{\prime}}\ell}^{\left(  2\right)  } &  \equiv\sum_{N,N^{\prime
}}P_{N}^{\left(  0\right)  }\omega_{\ell^{^{\prime}}N^{\prime},\ell N}%
^{2s}=\frac{2\pi}{\hbar}\sum_{N,N^{\prime}}P_{N}^{\left(  0\right)
}\left\vert H_{\ell N,\ell^{^{\prime}}N^{\prime}}^{^{\prime}}\right\vert
^{2}\delta\left(  E_{N}-E_{N^{\prime}}+\epsilon_{\ell}-\epsilon_{\ell
^{^{\prime}}}\right)  ,\label{scattering rate}\\
\omega_{\ell\ell^{^{\prime}}}^{\left(  2\right)  } &  =\sum_{N,N^{\prime}%
}P_{N}^{\left(  0\right)  }\omega_{\ell N^{\prime},\ell^{^{\prime}}N}%
^{2s}=\frac{2\pi}{\hbar}\sum_{N,N^{\prime}}P_{N}^{\left(  0\right)
}\left\vert H_{\ell N^{\prime},\ell^{^{\prime}}N}^{^{\prime}}\right\vert
^{2}\delta\left(  E_{N^{\prime}}-E_{N}+\epsilon_{\ell}-\epsilon_{\ell
^{^{\prime}}}\right)  .\nonumber
\end{align}
Now one has to introduce another basic statistical assumption, i.e.,%
\begin{equation}
\sum_{n}n_{\ell}n_{\ell^{^{\prime}}}\tilde{F}_{n}^{\left(  -2\right)
}=\left[  f_{\ell}f_{\ell^{^{\prime}}}\right]  ^{\left(  -2\right)  }\equiv
f_{\ell}^{0}f_{\ell^{^{\prime}}}^{\left(  -2\right)  }+f_{\ell}^{\left(
-2\right)  }f_{\ell^{^{\prime}}}^{0},\label{AMC}%
\end{equation}
which is analogous to the assumption of molecular chaos introduced in deriving
the classical Boltzmann equation from the classical Liouville equation (BBGKY
hierarchy) \cite{Kardar}. Therefore, under the assumptions
(\ref{statistical-1}) and (\ref{AMC}) one arrives at the Boltzmann equation
for $f_{\ell}^{\left(  -2\right)  }$:
\begin{equation}
e\mathbf{E}\cdot\frac{\partial f_{\ell}^{0}}{\hbar\partial\mathbf{k}}%
+\sum_{\ell^{^{\prime}}}\left[  \omega_{\ell^{^{\prime}}\ell}^{\left(
2\right)  }\left(  f_{\ell}^{\left(  -2\right)  }-\left[  f_{\ell}%
f_{\ell^{^{\prime}}}\right]  ^{\left(  -2\right)  }\right)  -\omega_{\ell
\ell^{^{\prime}}}^{\left(  2\right)  }\left(  f_{\ell^{^{\prime}}}^{\left(
-2\right)  }-\left[  f_{\ell}f_{\ell^{^{\prime}}}\right]  ^{\left(  -2\right)
}\right)  \right]  =0,
\end{equation}
which is just the linearized Bloch-Boltzmann equation. Utilizing the
microscopic detailed balance that can be verified directly in the lowest order
perturbation theory, one has
\begin{equation}
\omega_{\ell^{^{\prime}}\ell}^{\left(  2\right)  }f_{\ell}^{0}\left(
1-f_{\ell^{^{\prime}}}^{0}\right)  =\omega_{\ell\ell^{^{\prime}}}^{\left(
2\right)  }f_{\ell^{^{\prime}}}^{0}\left(  1-f_{\ell}^{0}\right)
\label{detailed balance}%
\end{equation}
and ($\delta f_{\ell}\equiv f_{\ell}-f_{\ell}^{0}$)
\begin{equation}
\delta f_{\ell}\left(  1-f_{\ell^{^{\prime}}}^{0}\right)  +f_{\ell}^{0}\left(
-\delta f_{\ell^{^{\prime}}}\right)  -\frac{f_{\ell}^{0}\left(  1-f_{\ell
^{^{\prime}}}^{0}\right)  }{f_{\ell^{^{\prime}}}^{0}\left(  1-f_{\ell}%
^{0}\right)  }\left[  \delta f_{\ell^{^{\prime}}}\left(  1-f_{\ell}%
^{0}\right)  +f_{\ell^{^{\prime}}}^{0}\left(  -\delta f_{\ell}\right)
\right]  =\delta f_{\ell}\frac{1-f_{\ell^{^{\prime}}}^{0}}{1-f_{\ell}^{0}%
}-\delta f_{\ell^{^{\prime}}}\frac{f_{\ell}^{0}}{f_{\ell^{^{\prime}}}^{0}%
},\label{linearization}%
\end{equation}
thus%
\begin{equation}
e\mathbf{E}\cdot\frac{\partial f_{\ell}^{0}}{\hbar\partial\mathbf{k}}%
+\sum_{\ell^{^{\prime}}}\omega_{\ell^{^{\prime}}\ell}^{\left(  2\right)
}\left[  f_{\ell}^{\left(  -2\right)  }\frac{1-f_{\ell^{^{\prime}}}^{0}%
}{1-f_{\ell}^{0}}-f_{\ell^{^{\prime}}}^{\left(  -2\right)  }\frac{f_{\ell}%
^{0}}{f_{\ell^{^{\prime}}}^{0}}\right]  =0,
\end{equation}
which is just the practical form of the Bloch-Boltzmann equation, i.e., Eq.
(\ref{SBE-n}) in the main text (note that $\omega_{\ell^{^{\prime}}\ell
}^{\left(  2\right)  }\equiv w_{\ell^{\prime}\ell}$ and $f_{\ell}^{\left(
-2\right)  }=\delta f_{\ell}^{n}$).

In the case of static disorder, the conventional skew scattering appears in
the Boltzmann equation in the first order of disorder potential
\cite{Sinitsyn2008}. The harmonic approximation is assumed for the scattering
system, then one has $\tilde{\omega}_{L^{^{\prime}}L}^{\left(  3\right)
}=\tilde{\omega}_{LL^{^{\prime}}}^{\left(  3\right)  }=0$, $\tilde{C}%
_{L}^{\left(  1\right)  }=0$ and $\tilde{C}_{LL^{^{\prime}}}^{\left(
0\right)  }\tilde{H}_{L^{^{\prime}}L}^{^{\prime}}=0$. Thus $\tilde{F}%
_{L}^{\left(  -1\right)  }=0$ and $f_{\ell}^{\left(  -1\right)  }=0$. This
leads to vanishing conventional skew scattering due to phonons, as pointed out
in Refs. \cite{Bruno2001,Hou2015,Lyo1973} and experimentally confirmed in
Refs. \cite{Hou2015,Tian2009}.

\subsection{Anomalous distribution function}

In the second order of disorder potential the transport equation for
$\tilde{F}_{L}^{\left(  0\right)  }$ can be decomposed into%
\begin{equation}
0=\tilde{C}_{L}^{^{\prime\prime}}+i\hbar\sum_{L^{^{\prime}}}\tilde{\omega
}_{LL^{^{\prime}}}^{\left(  2\right)  }\left[  \tilde{F}_{L}^{\left(
0\right)  ,a}-\tilde{F}_{L^{^{\prime}}}^{\left(  0\right)  ,a}\right]
\label{F0}%
\end{equation}
and $0=\sum_{L^{^{\prime}}}\tilde{\omega}_{LL^{^{\prime}}}^{\left(  2\right)
}\left[  \tilde{F}_{L}^{\left(  0\right)  ,n}-\tilde{F}_{L^{^{\prime}}%
}^{\left(  0\right)  ,n}\right]  +i\hbar\sum_{L^{^{\prime}}}\left[
\tilde{\omega}_{L^{^{\prime}}L}^{\left(  4\right)  }\tilde{F}_{L}^{\left(
-2\right)  }-\tilde{\omega}_{LL^{^{\prime}}}^{\left(  4\right)  }\tilde
{F}_{L^{^{\prime}}}^{\left(  -2\right)  }\right]  $, where $\tilde{F}%
_{L}^{\left(  0\right)  }=\tilde{F}_{L}^{\left(  0\right)  ,n}+\tilde{F}%
_{L}^{\left(  0\right)  ,a}$ and $\tilde{C}_{L}^{^{\prime\prime}}$ is given by
Eq. (\ref{anomalous driving}). Here we only analyze the equation for
$\tilde{F}_{L}^{\left(  0\right)  ,a}$, yielding the anomalous distribution
that is related to the side jump effect. $\tilde{F}_{L}^{\left(  0\right)
,n}$ is related to the so-called intrinsic skew scattering, which is not
likely to have an intuitive generic description in the case of dynamical
disorder \cite{Lyo1973}.

Utilizing$\ $%
\[
\sum_{nN}n_{k}\sum_{n^{\prime}N^{\prime}}^{\prime}\left[  \tilde
{C}_{nN,n^{\prime}N^{\prime}}^{\left(  1\right)  }\tilde{H}_{n^{\prime
}N^{\prime},nN}^{^{\prime}}/d_{nN,n^{\prime}N^{\prime}}^{-}-c.c.\right]
=\sum_{nN,n^{\prime}N^{\prime}}^{\prime}\left(  n_{k}-n_{k}^{\prime}\right)
\tilde{C}_{nN,n^{\prime}N^{\prime}}^{\left(  1\right)  }\tilde{H}_{n^{\prime
}N^{\prime},nN}^{^{\prime}}/d_{nN,n^{\prime}N^{\prime}}^{-}%
\]
and Eq. (\ref{operator-2}) and similar techniques to those in deriving the
conventional Bloch-Boltzmann equation, we get%
\begin{align*}
\sum_{nN}n_{\ell}\tilde{C}_{nN}^{^{\prime\prime}} &  =-\sum_{\ell^{^{\prime}}%
}^{\prime}\sum_{nN,N^{\prime}}^{\prime}\frac{1}{2}\left(  -\beta\right)
i\hbar e\mathbf{E\cdot}\left[  i\mathbf{J}_{\ell^{^{\prime}}}-i\mathbf{J}%
_{\ell}+\mathbf{\hat{D}}\arg H_{\ell N,\ell^{^{\prime}}N^{\prime}}^{^{\prime}%
}\right]  \omega_{\ell^{^{\prime}}N^{\prime},\ell N}^{2s}n_{\ell}\left(
1-n_{\ell^{^{\prime}}}\right)  \tilde{\rho}_{nN}^{\left(  0\right)  }\\
&  -\sum_{\ell^{^{\prime}}}^{\prime}\sum_{nN,N^{\prime}}^{\prime}\frac{1}%
{2}\left(  -\beta\right)  i\hbar e\mathbf{E\cdot}\left[  i\mathbf{J}%
_{\ell^{^{\prime}}}-i\mathbf{J}_{\ell}+\mathbf{\hat{D}}\arg H_{\ell N^{\prime
},\ell^{^{\prime}}N}^{^{\prime}}\right]  \omega_{\ell N^{\prime}%
,\ell^{^{\prime}}N}^{2s}n_{\ell^{^{\prime}}}\left(  1-n_{\ell}\right)
\tilde{\rho}_{nN}^{\left(  0\right)  }.
\end{align*}
Notice $\arg H_{\ell N,\ell^{^{\prime}}N^{\prime}}^{^{\prime}}{}=\arg H_{\ell
N^{\prime},\ell^{^{\prime}}N}^{^{\prime}}=\arg H_{\ell\ell^{^{\prime}}%
}^{^{\prime}}{}$ because the the quanta of the scattering system is boson, and
$\tilde{\rho}_{nN}^{\left(  0\right)  }=P_{N}^{\left(  0\right)  }\tilde
{F}_{n}^{\left(  0\right)  }$, we obtain%
\begin{equation}
\sum_{nN}n_{\ell}\tilde{C}_{nN}^{^{\prime\prime}}=-\frac{1}{2}\left(
-\beta\right)  i\hbar e\mathbf{E\cdot}\sum_{\ell^{^{\prime}}}^{\prime}%
\delta\mathbf{r}_{\ell^{^{\prime}}\ell}\left[  \omega_{\ell^{^{\prime}}\ell
}^{\left(  2\right)  }f_{\ell}^{0}\left(  1-f_{\ell^{^{\prime}}}^{0}\right)
+\omega_{\ell\ell^{^{\prime}}}^{\left(  2\right)  }f_{\ell^{^{\prime}}}%
^{0}\left(  1-f_{\ell}^{0}\right)  \right]  .
\end{equation}
By Eq. (\ref{detailed balance}), we obtain%
\begin{equation}
\sum_{nN}n_{\ell}\tilde{C}_{nN}^{^{\prime\prime}}=-\left(  -\beta\right)
i\hbar e\mathbf{E\cdot}\sum_{\ell^{^{\prime}}}\delta\mathbf{r}_{\ell
^{^{\prime}}\ell}\omega_{\ell^{^{\prime}}\ell}^{\left(  2\right)  }f_{\ell
}^{0}\left(  1-f_{\ell^{^{\prime}}}^{0}\right)  =-i\hbar e\mathbf{E\cdot
}\left[  \sum_{\ell^{^{\prime}}}\frac{1-f_{\ell^{^{\prime}}}^{0}}{1-f_{\ell
}^{0}}\omega_{\ell^{^{\prime}}\ell}^{\left(  2\right)  }\delta\mathbf{r}%
_{\ell^{^{\prime}}\ell}\right]  \frac{\partial f_{\ell}^{0}}{\partial
\epsilon_{\ell}}.
\end{equation}
Then we treat the collision term by employing the basic assumption
\begin{equation}
\tilde{F}_{nN}^{\left(  0\right)  ,a}=P_{N}^{\left(  0\right)  }\tilde{F}%
_{n}^{\left(  0\right)  ,a}\label{statistical-a}%
\end{equation}
and the \textquotedblleft assumption of molecular chaos\textquotedblright%
\begin{equation}
\sum_{n}n_{\ell}n_{\ell^{^{\prime}}}\tilde{F}_{n}^{\left(  0\right)
,a}=\left[  f_{\ell}f_{\ell^{^{\prime}}}\right]  ^{\left(  0\right)  ,a}\equiv
f_{\ell}^{\left(  0\right)  ,a}f_{\ell^{^{\prime}}}^{0}+f_{\ell}^{0}%
f_{\ell^{^{\prime}}}^{\left(  0\right)  ,a},\label{AMC-a}%
\end{equation}
yielding the Boltzmann equation for $f_{\ell}^{\left(  0\right)  ,a}$:
\begin{equation}
0=-e\mathbf{E\cdot}\left[  \sum_{\ell^{^{\prime}}}\frac{1-f_{\ell^{^{\prime}}%
}^{0}}{1-f_{\ell}^{0}}\omega_{\ell^{^{\prime}}\ell}^{\left(  2\right)  }%
\delta\mathbf{r}_{\ell^{^{\prime}}\ell}\right]  \frac{\partial f_{\ell}^{0}%
}{\partial\epsilon_{\ell}}+\sum_{\ell^{^{\prime}}}\left[  \omega
_{\ell^{^{\prime}}\ell}^{\left(  2\right)  }\left[  f_{\ell}\left(
1-f_{\ell^{^{\prime}}}\right)  \right]  ^{\left(  0\right)  ,a}-\omega
_{\ell\ell^{^{\prime}}}^{\left(  2\right)  }\left[  f_{\ell^{^{\prime}}%
}\left(  1-f_{\ell}\right)  \right]  ^{\left(  0\right)  ,a}\right]  .
\end{equation}
Utilizing Eqs. (\ref{detailed balance}) and (\ref{linearization}), we get%
\begin{equation}
0=-e\mathbf{E\cdot}\left[  \sum_{\ell^{^{\prime}}}\frac{1-f_{\ell^{^{\prime}}%
}^{0}}{1-f_{\ell}^{0}}\omega_{\ell^{^{\prime}}\ell}^{\left(  2\right)  }%
\delta\mathbf{r}_{\ell^{^{\prime}}\ell}\right]  \frac{\partial f_{\ell}^{0}%
}{\partial\epsilon_{\ell}}+\sum_{\ell^{^{\prime}}}\omega_{\ell^{^{\prime}}%
\ell}^{\left(  2\right)  }\left[  f_{\ell}^{\left(  0\right)  ,a}%
\frac{1-f_{\ell^{^{\prime}}}^{0}}{1-f_{\ell}^{0}}-f_{\ell^{^{\prime}}%
}^{\left(  0\right)  ,a}\frac{f_{\ell}^{0}}{f_{\ell^{^{\prime}}}^{0}}\right]
.
\end{equation}
This is exactly the same Boltzmann equation for the anomalous distribution
function $f_{\ell}^{\left(  0\right)  ,a}\equiv\delta f_{\ell}^{a}$ as we
obtained via phenomenological arguments in the main text.

\subsection{Berry curvature anomalous velocity and side-jump velocity}

For the observables of interest, $\tilde{A}$ is diagonal with respect to $N$,
hence $\tilde{F}_{LL^{^{\prime}}}^{\left(  -1\right)  }$ does not contribute
to the off-diagonal response, and the off-diagonal response $\sum
_{LL^{^{\prime}}}^{^{\prime}}\tilde{F}_{LL^{^{\prime}}}\tilde{A}_{L^{^{\prime
}}L}$ is equal to%
\begin{equation}
\sum_{LL^{^{\prime}}}^{\prime}\tilde{F}_{LL^{^{\prime}}}^{\left(  0\right)
}\tilde{A}_{L^{^{\prime}}L}=\delta^{\text{in}}A+\delta^{\text{sj}}A,
\end{equation}
where%
\begin{equation}
\delta^{\text{in}}A\equiv\sum_{LL^{^{\prime}}}^{\prime}C_{LL^{^{\prime}}%
}^{\left(  0\right)  }\frac{\tilde{A}_{L^{^{\prime}}L}}{E_{L}-E_{L^{^{\prime}%
}}-i\hbar s}%
\end{equation}
is the intrinsic part, whereas
\begin{equation}
\delta^{\text{sj}}A\equiv\sum_{LL^{^{\prime}}L^{^{\prime\prime}}}^{\prime
}\tilde{F}_{L}^{\left(  -2\right)  }\left[  \left(  \frac{\tilde
{H}_{L^{^{\prime}}L^{^{\prime\prime}}}^{^{\prime}}\tilde{H}_{L^{^{\prime
\prime}}L}^{^{\prime}}\tilde{A}_{LL^{^{\prime}}}}{d_{LL^{^{\prime\prime}}}%
^{+}d_{LL^{^{\prime}}}^{+}}+c.c.\right)  +\frac{\tilde{H}_{LL^{^{\prime}}%
}^{^{\prime}}\tilde{H}_{L^{^{\prime\prime}}L}^{^{\prime}}\tilde{A}%
_{L^{^{\prime}}L^{^{\prime\prime}}}}{d_{LL^{^{\prime\prime}}}^{+}%
d_{LL^{^{\prime}}}^{-}}\right]
\end{equation}
is the disorder-dependent part.

\subsubsection{Intrinsic contribution: electric-field induced
interband-coherence}

Due to Eqs. (\ref{C-0}) and (\ref{operator-2}), we have ($\tilde{\rho}%
_{nN}^{\left(  0\right)  }=P_{N}^{\left(  0\right)  }\tilde{\rho}_{n}^{\left(
0\right)  }$)%
\begin{align*}
\delta^{\text{in}}A &  =\sum_{n,n^{\prime}}^{\prime}\sum_{N}ie\mathbf{E}%
\cdot\sum_{\ell\ell^{^{\prime}}}^{\prime}\mathbf{J}_{\ell\ell^{^{\prime}}%
}\left(  e^{-\beta\left(  \epsilon_{\ell^{^{\prime}}}-\epsilon_{\ell}\right)
}-1\right)  \tilde{\rho}_{nN}^{\left(  0\right)  }\left(  a_{\ell}^{\dag
}a_{\ell^{^{\prime}}}\right)  _{n,n^{\prime}}\frac{\tilde{A}_{n^{\prime}N,nN}%
}{E_{n}-E_{n^{\prime}}-i\hbar s}\\
&  =\sum_{n,n^{\prime}}^{\prime}ie\mathbf{E}\cdot\sum_{\ell\ell^{^{\prime}}%
}^{\prime}\mathbf{J}_{\ell\ell^{^{\prime}}}A_{\ell^{^{\prime}}\ell}%
\frac{-\tilde{\rho}_{n}^{\left(  0\right)  }}{\epsilon_{\ell}-\epsilon
_{\ell^{^{\prime}}}-i\hbar s}\left[  n_{\ell}\left(  1-n_{\ell^{^{\prime}}%
}\right)  \delta_{n_{\ell}-1=n_{\ell}^{\prime}}\delta_{n_{\ell^{^{\prime}}%
}+1=n_{\ell^{^{\prime}}}^{\prime}}-n_{\ell}^{\prime}\left(  1-n_{\ell
^{^{\prime}}}^{\prime}\right)  \delta_{n_{\ell}^{\prime}-1=n_{\ell}}%
\delta_{n_{\ell^{^{\prime}}}^{\prime}+1=n_{\ell^{^{\prime}}}}\right]  ,
\end{align*}
where we used $\tilde{\rho}_{n}^{\left(  0\right)  }\left[  e^{-\beta\left(
E_{n^{\prime}}-E_{n}\right)  }-1\right]  =\tilde{\rho}_{n^{\prime}}^{\left(
0\right)  }-\tilde{\rho}_{n}^{\left(  0\right)  }$. Notice that for fermions%
\[
n_{\ell}^{\prime}\left(  1-n_{\ell^{^{\prime}}}^{\prime}\right)
\delta_{n_{\ell}^{\prime}-1=n_{\ell}}\delta_{n_{\ell^{^{\prime}}}^{\prime
}+1=n_{\ell^{^{\prime}}}}=\left(  1-n_{\ell}\right)  n_{\ell^{^{\prime}}%
}\delta_{n_{\ell}^{\prime}-1=n_{\ell}}\delta_{n_{\ell^{^{\prime}}}^{\prime
}+1=n_{\ell^{^{\prime}}}},
\]
we get
\begin{equation}
\delta^{\text{in}}A=-ie\mathbf{E}\cdot\sum_{n}\sum_{\ell\ell^{^{\prime}}%
}^{\prime}\mathbf{J}_{\ell\ell^{^{\prime}}}A_{\ell^{^{\prime}}\ell}%
\frac{\tilde{\rho}_{n}^{\left(  0\right)  }}{\epsilon_{\ell}-\epsilon
_{\ell^{^{\prime}}}-i\hbar s}\left[  n_{\ell}\left(  1-n_{\ell^{^{\prime}}%
}\right)  -n_{\ell^{^{\prime}}}\left(  1-n_{\ell}\right)  \right]  =\sum
_{\ell\ell^{^{\prime}}}^{\prime}\frac{C_{\ell\ell^{^{\prime}}}^{\left(
0\right)  }A_{\ell^{^{\prime}}\ell}}{\epsilon_{\ell}-\epsilon_{\ell^{^{\prime
}}}-i\hbar s},
\end{equation}
where $C_{\ell\ell^{\prime}}^{\left(  0\right)  }=ie\mathbf{E\cdot J}%
_{\ell\ell^{\prime}}\left(  f_{\ell^{\prime}}^{0}-f_{\ell}^{0}\right)  $. This
is just the intrinsic contribution $\delta^{\text{in}}A\equiv\sum_{\ell
}f_{\ell}^{0}\delta^{\text{in}}A_{\ell}$ to linear response with respect to
the uniform and time-independent electric field. Here we use $\mathbf{v}%
_{\ell\ell^{^{\prime}}}\delta_{\mathbf{kk}^{^{\prime}}}=-\frac{1}{\hbar
}\left(  \epsilon_{\ell}-\epsilon_{\ell^{^{\prime}}}\right)  \mathbf{J}%
_{\ell\ell^{^{\prime}}}$ for $\ell\neq\ell^{^{\prime}}$, and $\delta
^{\text{in}}A_{\ell}$ is just the intrinsic correction to $A_{\ell}$ in the
semiclassical Boltzmann formulation \cite{Xiao2017SOT-SBE}. In the case of
$A=\mathbf{j}=e\mathbf{v}$, $\delta^{\text{in}}\mathbf{v}_{\ell}%
=\mathbf{v}_{\ell}^{\text{bc}}$ is the Berry-curvature anomalous velocity.

\subsubsection{Side-jump velocity: scattering-induced interband-coherence}

Now we analyze $\delta^{\text{sj}}A$. Here%
\begin{align*}
&  \sum_{nN,n^{\prime}N^{\prime},n^{\prime\prime}N^{\prime\prime}}^{\prime
}\tilde{F}_{nN}^{\left(  -2\right)  }\frac{\tilde{H}_{nN,n^{\prime}N^{\prime}%
}^{^{\prime}}\tilde{H}_{n^{\prime\prime}N^{\prime\prime},nN}^{^{\prime}}%
\tilde{A}_{n^{\prime},n^{\prime\prime}}}{\left(  E_{nN}-E_{n^{\prime}%
N^{\prime}}-i\hbar s\right)  \left(  E_{nN}-E_{n^{\prime\prime}N^{\prime
\prime}}+i\hbar s\right)  }\\
&  =\sum_{n,n^{\prime},n^{\prime\prime}}^{\prime}\sum_{N,N^{\prime}}\tilde
{F}_{nN}^{\left(  -2\right)  }\frac{\sum_{\ell\ell^{^{\prime}}}^{^{\prime}%
}\sum_{kk^{\prime}}^{^{\prime}}\sum_{jj^{\prime}}^{^{\prime}}H_{\ell
N,\ell^{^{\prime}}N^{\prime}}^{^{\prime}}H_{k^{\prime}N^{\prime},kN}%
^{^{\prime}}A_{j^{\prime}j}\left(  a_{\ell}^{\dag}a_{\ell^{^{\prime}}}\right)
_{n,n^{\prime}}\left(  a_{j^{\prime}}^{\dag}a_{j}\right)  _{n^{\prime
},n^{\prime\prime}}\left(  a_{k^{\prime}}^{\dag}a_{k}\right)  _{n^{\prime
\prime},n}}{\left(  E_{nN}-E_{n^{\prime}N^{\prime}}-i\hbar s\right)  \left(
E_{nN}-E_{n^{\prime\prime}N^{\prime}}+i\hbar s\right)  },
\end{align*}
since $N^{^{\prime}}=N^{^{\prime\prime}}$ and then $n^{^{\prime}}\neq
n^{^{\prime\prime}}$ and thus $j\neq j^{^{\prime}}$. Using%
\begin{align*}
&  \left(  a_{\ell}^{\dag}a_{\ell^{^{\prime}}}\right)  _{n,n^{\prime}}\left(
a_{j^{\prime}}^{\dag}a_{j}\right)  _{n^{\prime},n^{\prime\prime}}\left(
a_{k^{\prime}}^{\dag}a_{k}\right)  _{n^{\prime\prime},n}=\delta_{k^{\prime}%
j}\delta_{j^{\prime}\ell^{^{\prime}}}\delta_{k\ell}n_{\ell}\left(
1-n_{j}\right)  \left(  1-n_{j^{\prime}}\right)  \delta_{n_{k}-1=n_{k}%
^{^{\prime\prime}}}\delta_{n_{j}+1=n_{j}^{^{\prime\prime}}}\delta
_{n_{j^{^{\prime}}}^{^{\prime\prime}}+1=n_{j^{^{\prime}}}^{^{\prime}}}%
\delta_{n_{j},n_{j}^{^{\prime}}}\delta_{n_{j^{^{\prime}}}^{^{\prime\prime}%
},n_{j^{^{\prime}}}}\delta_{n_{k}^{^{\prime}},n_{k}^{^{\prime\prime}}}\\
&  -\delta_{kj^{\prime}}\delta_{k^{\prime}\ell^{^{\prime}}}\delta_{j\ell
}n_{j^{\prime}}\left(  1-n_{\ell^{^{\prime}}}\right)  n_{j}\delta
_{n_{j^{^{\prime}}}-1=n_{j^{^{\prime}}}^{^{\prime\prime}}}\delta
_{n_{\ell^{^{\prime}}}+1=n_{\ell^{^{\prime}}}^{^{\prime\prime}}}\delta
_{n_{j}^{^{\prime\prime}}-1=n_{j}^{^{\prime}}}\delta_{n_{j^{^{\prime}}%
}=n_{j^{^{\prime}}}^{^{\prime}}}\delta_{n_{\ell^{^{\prime}}}^{^{\prime}%
}=n_{\ell^{^{\prime}}}^{^{\prime\prime}}}\delta_{n_{j},n_{j}^{^{\prime\prime}%
}},
\end{align*}
we get%
\begin{align*}
&  \sum_{nN,n^{\prime}N^{\prime},n^{\prime\prime}N^{\prime\prime}}^{\prime
}\tilde{F}_{nN}^{\left(  -2\right)  }\frac{\tilde{H}_{nN,n^{\prime}N^{\prime}%
}^{^{\prime}}\tilde{H}_{n^{\prime\prime}N^{\prime\prime},nN}^{^{\prime}}%
\tilde{A}_{n^{\prime},n^{\prime\prime}}}{\left(  E_{nN}-E_{n^{\prime}%
N^{\prime}}-i\hbar s\right)  \left(  E_{nN}-E_{n^{\prime\prime}N^{\prime
\prime}}+i\hbar s\right)  }\\
&  =\sum_{n}\sum_{N,N^{\prime}}\tilde{F}_{n}^{\left(  -2\right)  }\frac
{\sum_{\ell jj^{\prime}}^{^{\prime}}P_{N}^{\left(  0\right)  }H_{\ell
N,j^{\prime}N^{\prime}}^{^{\prime}}H_{jN^{\prime},\ell N}^{^{\prime}%
}A_{j^{\prime}j}n_{\ell}\left(  1-n_{j}\right)  \left(  1-n_{j^{\prime}%
}\right)  }{\left(  E_{N}-E_{N^{\prime}}+\epsilon_{\ell}-\epsilon_{j^{\prime}%
}-i\hbar s\right)  \left(  E_{N}-E_{N^{\prime}}+\epsilon_{\ell}-\epsilon
_{j}+i\hbar s\right)  }\\
&  -\sum_{n}\sum_{N,N^{\prime}}\tilde{F}_{n}^{\left(  -2\right)  }\frac
{\sum_{\ell jj^{\prime}}^{^{\prime}}P_{N^{\prime}}^{\left(  0\right)  }H_{\ell
N,j^{\prime}N^{\prime}}^{^{\prime}}H_{jN^{\prime},\ell N}^{^{\prime}%
}A_{j^{\prime}j}\left(  1-n_{\ell}\right)  n_{j}n_{j^{\prime}}}{\left(
E_{N}-E_{N^{\prime}}+\epsilon_{\ell}-\epsilon_{j^{\prime}}-i\hbar s\right)
\left(  E_{N}-E_{N^{\prime}}+\epsilon_{\ell}-\epsilon_{j}+i\hbar s\right)  },
\end{align*}
where we have applied the assumption (\ref{statistical-1}). In the case of
$A=\mathbf{v}$, $v_{j^{\prime}j}=\frac{1}{i\hbar}r_{j^{\prime}j}\left(
\epsilon_{j}-\epsilon_{j^{\prime}}\right)  $ thus%
\begin{align*}
&  \sum_{nN,n^{\prime}N^{\prime},n^{\prime\prime}N^{\prime\prime}}^{\prime
}\tilde{F}_{nN}^{\left(  -2\right)  }\frac{\tilde{H}_{nN,n^{\prime}N^{\prime}%
}^{^{\prime}}\tilde{H}_{n^{\prime\prime}N^{\prime\prime},nN}^{^{\prime}}%
\tilde{A}_{n^{\prime},n^{\prime\prime}}}{\left(  E_{nN}-E_{n^{\prime}%
N^{\prime}}-i\hbar s\right)  \left(  E_{nN}-E_{n^{\prime\prime}N^{\prime
\prime}}+i\hbar s\right)  }\\
&  =2\operatorname{Re}\sum_{\ell jj^{\prime}}^{\prime}\frac{i}{\hbar}\sum
_{n}\tilde{F}_{n}^{\left(  -2\right)  }n_{\ell}\left(  1-n_{j}\right)
\sum_{N,N^{\prime}}P_{N}^{\left(  0\right)  }\frac{H_{\ell N,jN^{\prime}%
}^{^{\prime}}r_{jj^{\prime}}H_{j^{\prime}N^{\prime},\ell N}^{^{\prime}}}%
{E_{N}-E_{N^{\prime}}+\epsilon_{\ell}-\epsilon_{j}-i\hbar s}\\
&  -2\operatorname{Re}\sum_{\ell jj^{\prime}}^{\prime}\frac{i}{\hbar}\sum
_{n}\tilde{F}_{n}^{\left(  -2\right)  }n_{\ell}\left(  1-n_{j}\right)
n_{j^{\prime}}\sum_{N,N^{\prime}}P_{N}^{\left(  0\right)  }\frac{H_{\ell
N,jN^{\prime}}^{^{\prime}}r_{jj^{\prime}}H_{j^{\prime}N^{\prime},\ell
N}^{^{\prime}}}{E_{N}-E_{N^{\prime}}+\epsilon_{\ell}-\epsilon_{j}-i\hbar s}\\
&  -2\operatorname{Re}\sum_{\ell jj^{\prime}}^{\prime}\frac{i}{\hbar}\sum
_{n}\tilde{F}_{n}^{\left(  -2\right)  }\left(  1-n_{j}\right)  n_{\ell
}n_{j^{\prime}}\sum_{N,N^{\prime}}P_{N}^{\left(  0\right)  }\frac
{H_{jN^{\prime},j^{\prime}N}^{^{\prime}}r_{j^{\prime}\ell}H_{\ell
N,jN^{\prime}}^{^{\prime}}}{E_{N}-E_{N^{\prime}}+\epsilon_{\ell}-\epsilon
_{j}-i\hbar s}.
\end{align*}
%where the last term is obtained via%
%\begin{align*}
%&  -2\operatorname{Re}\sum_{\ell jj^{\prime}}\ '\frac{1}{i\hbar}%
%\sum_{n}\tilde{F}_{n}^{\left(  -2\right)  }\left(  1-n_{\ell}\right)
%n_{j}n_{j^{\prime}}\sum_{N,N^{\prime}}\ 'P_{N^{\prime}}^{\left(
%0\right)  }\frac{H_{\ell N,j^{\prime}N^{\prime}}^{^{\prime}}r_{j^{\prime}%
%j}H_{jN^{\prime},\ell N}^{^{\prime}}}{E_{N}-E_{N^{\prime}}+\epsilon_{\ell
%}-\epsilon_{j}+i\hbar s}\\
%&  =-2\operatorname{Re}\sum_{\ell jj^{\prime}}\ '\frac{1}{i\hbar}%
%\sum_{n}\tilde{F}_{n}^{\left(  -2\right)  }\left(  1-n_{j}\right)  n_{\ell
%}n_{j^{\prime}}\sum_{N,N^{\prime}}\ 'P_{N}^{\left(  0\right)  }%
%\frac{H_{jN^{\prime},j^{\prime}N}^{^{\prime}}r_{j^{\prime}\ell}H_{\ell
%N,jN^{\prime}}^{^{\prime}}}{E_{N^{\prime}}-E_{N}+\epsilon_{j}-\epsilon_{\ell
%}+i\hbar s}.
%\end{align*}
The reason for writing the last term in this form will be clear soon. Thus we
get%
\begin{align}
&  \sum_{nN,n^{\prime}N^{\prime},n^{\prime\prime}N^{\prime\prime}}^{\prime
}\tilde{F}_{nN}^{\left(  -2\right)  }\frac{\tilde{H}_{nN,n^{\prime}N^{\prime}%
}^{^{\prime}}\tilde{H}_{n^{\prime\prime}N^{\prime\prime},nN}^{^{\prime}}%
\tilde{A}_{n^{\prime},n^{\prime\prime}}}{\left(  E_{nN}-E_{n^{\prime}%
N^{\prime}}-i\hbar s\right)  \left(  E_{nN}-E_{n^{\prime\prime}N^{\prime
\prime}}+i\hbar s\right)  }\label{KLA-1}\\
&  =2\operatorname{Re}\sum_{\ell jj^{\prime}}^{\prime}\frac{i}{\hbar}\sum
_{n}\tilde{F}_{n}^{\left(  -2\right)  }n_{\ell}\left(  1-n_{j}\right)
\sum_{N,N^{\prime}}P_{N}^{\left(  0\right)  }\frac{H_{\ell N,jN^{\prime}%
}^{^{\prime}}r_{jj^{\prime}}H_{j^{\prime}N^{\prime},\ell N}^{^{\prime}}}%
{E_{N}-E_{N^{\prime}}+\epsilon_{\ell}-\epsilon_{j}-i\hbar s}\nonumber\\
&  -2\operatorname{Re}\sum_{\ell jj^{\prime}}^{\prime}\frac{i}{\hbar}\sum
_{n}\tilde{F}_{n}^{\left(  -2\right)  }n_{\ell}\left(  1-n_{j}\right)
n_{j^{\prime}}\sum_{N,N^{\prime}}P_{N}^{\left(  0\right)  }\frac{H_{\ell
N,jN^{\prime}}^{^{\prime}}\left[  r_{jj^{\prime}}H_{j^{\prime}N^{\prime},\ell
N}^{^{\prime}}+H_{jN^{\prime},j^{\prime}N}^{^{\prime}}r_{j^{\prime}\ell
}\right]  }{E_{N}-E_{N^{\prime}}+\epsilon_{\ell}-\epsilon_{j}-i\hbar
s}.\nonumber
\end{align}
Besides, we have%
\begin{align*}
&  \sum_{nN,n^{\prime}N^{\prime},n^{\prime\prime}N^{\prime\prime}}^{\prime
}\tilde{F}_{nN}^{\left(  -2\right)  }\left[  \frac{\tilde{H}_{n^{\prime
}N^{\prime},n^{\prime\prime}N^{\prime\prime}}^{^{\prime}}\tilde{H}%
_{n^{\prime\prime}N^{\prime\prime},nN}^{^{\prime}}\tilde{A}_{n,n^{\prime}}%
}{\left(  E_{nN}-E_{n^{\prime}N^{\prime}}+i\hbar s\right)  \left(
E_{nN}-E_{n^{\prime\prime}N^{\prime\prime}}+i\hbar s\right)  }+c.c.\right]  \\
&  =\sum_{n}\sum_{N,N^{\prime}}\tilde{F}_{nN}^{\left(  -2\right)  }\left[
\frac{\sum_{\ell jj^{\prime}}^{^{\prime}}H_{jN,j^{\prime}N^{\prime}}%
^{^{\prime}}H_{j^{\prime}N^{\prime},\ell N}^{^{\prime}}A_{\ell j}n_{\ell
}\left(  1-n_{j^{\prime}}\right)  \left(  1-n_{j}\right)  }{\left(
\epsilon_{\ell}-\epsilon_{j}+i\hbar s\right)  \left(  E_{N}-E_{N^{\prime}%
}+\epsilon_{\ell}-\epsilon_{j^{\prime}}+i\hbar s\right)  }+c.c.\right]  \\
&  -\sum_{n}\sum_{N,N^{\prime}}\tilde{F}_{nN}^{\left(  -2\right)  }\left[
\frac{\sum_{\ell jj^{\prime}}^{^{\prime}}H_{\ell N,j^{\prime}N^{\prime}%
}^{^{\prime}}H_{jN^{\prime},\ell N}^{^{\prime}}A_{j^{\prime}j}n_{\ell}\left(
1-n_{j}\right)  n_{j^{\prime}}}{\left(  \epsilon_{j^{\prime}}-\epsilon
_{j}+i\hbar s\right)  \left(  E_{N}-E_{N^{\prime}}+\epsilon_{\ell}%
-\epsilon_{j}+i\hbar s\right)  }+c.c.\right]  .
\end{align*}
In the case of $A=\mathbf{v}$, $v_{j^{\prime}j}=\frac{1}{i\hbar}r_{j^{\prime
}j}\left(  \epsilon_{j}-\epsilon_{j^{\prime}}\right)  $ thus%
\begin{align*}
&  \sum_{LL^{^{\prime}}L^{^{\prime\prime}}}^{\prime}\tilde{F}_{L}^{\left(
-2\right)  }\left[  \frac{\tilde{H}_{L^{^{\prime}}L^{^{\prime\prime}}%
}^{^{\prime}}\tilde{H}_{L^{^{\prime\prime}}L}^{^{\prime}}\tilde{A}%
_{LL^{^{\prime}}}}{\left(  E_{L}-E_{L^{^{\prime\prime}}}+i\hbar s\right)
\left(  E_{L}-E_{L^{^{\prime}}}+i\hbar s\right)  }+c.c.\right]  \\
&  =-2\operatorname{Re}\sum_{\ell jj^{\prime}}^{\prime}\frac{i}{\hbar}\sum
_{n}\tilde{F}_{n}^{\left(  -2\right)  }n_{\ell}\left(  1-n_{j}\right)
\sum_{N,N^{\prime}}P_{N}^{\left(  0\right)  }\frac{H_{\ell N,jN^{\prime}%
}^{^{\prime}}H_{jN^{\prime},j^{\prime}N,}^{^{\prime}}r_{j^{\prime}\ell}}%
{E_{N}-E_{N^{\prime}}+\epsilon_{\ell}-\epsilon_{j}-i\hbar s}\\
&  +2\operatorname{Re}\sum_{\ell jj^{\prime}}^{\prime}\frac{i}{\hbar}\sum
_{n}\tilde{F}_{n}^{\left(  -2\right)  }n_{\ell}\left(  1-n_{j}\right)
n_{j^{\prime}}\sum_{N,N^{\prime}}P_{N}^{\left(  0\right)  }\frac{H_{\ell
N,jN^{\prime}}^{^{\prime}}\left[  H_{jN^{\prime},j^{\prime}N,}^{^{\prime}%
}r_{j^{\prime}\ell}+r_{jj^{\prime}}H_{j^{\prime}N^{\prime},\ell N}^{^{\prime}%
}\right]  }{E_{N}-E_{N^{\prime}}+\epsilon_{\ell}-\epsilon_{j}-i\hbar s}.
\end{align*}
Together with Eq. (\ref{KLA-1}), we obtain (the $\mathbf{D}\left\vert
H_{jN^{\prime},\ell N}^{^{\prime}}\right\vert ^{2}$ term is neglected as
trivial renormalization effect, as in Ref. \cite{Xiao2017SOT-SBE})%
\begin{align*}
&  \sum_{LL^{\prime}L^{\prime\prime}}^{\prime}\tilde{F}_{L}^{\left(
-2\right)  }\left[  \frac{\tilde{H}_{L^{\prime}L^{\prime\prime}}^{^{\prime}%
}\tilde{H}_{L^{\prime\prime}L}^{^{\prime}}\tilde{A}_{LL^{\prime}}}{\left(
E_{L}-E_{L^{\prime\prime}}+i\hbar s\right)  \left(  E_{L}-E_{L^{\prime}%
}+i\hbar s\right)  }+c.c.\right]  +\sum_{LL^{\prime}L^{\prime\prime}}^{\prime
}\tilde{F}_{L}^{\left(  -2\right)  }\frac{\tilde{H}_{LL^{\prime}}^{^{\prime}%
}\tilde{H}_{L^{\prime\prime}L}^{^{\prime}}\tilde{A}_{L^{\prime}L^{\prime
\prime}}}{\left(  E_{L}-E_{L^{\prime}}-i\hbar s\right)  \left(  E_{L}%
-E_{L^{\prime\prime}}+i\hbar s\right)  }\\
&  =-2\operatorname{Re}\frac{i}{\hbar}\sum_{\ell j}\sum_{n}\tilde{F}%
_{n}^{\left(  -2\right)  }n_{\ell}\left(  1-n_{j}\right)  \sum_{N,N^{\prime}%
}\frac{P_{N}^{\left(  0\right)  }}{E_{N}-E_{N^{\prime}}+\epsilon_{\ell
}-\epsilon_{j}-i\hbar s}\left\vert H_{\ell N,jN^{\prime}}^{^{\prime}%
}\right\vert ^{2}\left[  -\mathbf{D}\arg H_{jN^{\prime},\ell N}^{^{\prime}%
}-\left(  i\mathbf{J}_{\ell}-i\mathbf{J}_{j}\right)  \right]  ,
\end{align*}
which is equal to
\begin{align*}
&  \sum_{\ell j}\sum_{n}\tilde{F}_{n}^{\left(  -2\right)  }n_{\ell}\left(
1-n_{j}\right)  \sum_{N,N^{\prime}}P_{N}^{\left(  0\right)  }\frac{2\pi}%
{\hbar}\left\vert H_{\ell N,jN^{\prime}}^{^{\prime}}\right\vert ^{2}%
\delta\left(  E_{N}-E_{N^{\prime}}+\epsilon_{\ell}-\epsilon_{j}\right)
\left[  i\mathbf{J}_{j}-i\mathbf{J}_{\ell}-\mathbf{D}\arg H_{j,\ell}%
^{^{\prime}}\right]  \\
&  =\sum_{\ell}f_{\ell}^{\left(  -2\right)  }\left[  \sum_{\ell^{^{\prime}}%
}\omega_{\ell^{^{\prime}}\ell}^{\left(  2\right)  }\frac{1-f_{\ell^{^{\prime}%
}}^{0}}{1-f_{\ell}^{0}}\delta\mathbf{r}_{\ell^{^{\prime}}\ell}\right]  .
\end{align*}
Here we used $\omega_{\ell^{^{\prime}}\ell}^{\left(  2\right)  }\equiv
\sum_{N,N^{\prime}}P_{N}^{\left(  0\right)  }\omega_{\ell^{^{\prime}}%
N^{\prime},\ell N}^{2s}=\frac{2\pi}{\hbar}\sum_{N,N^{\prime}}P_{N}^{\left(
0\right)  }\left\vert H_{\ell N,\ell^{^{\prime}}N^{\prime}}^{^{\prime}%
}\right\vert ^{2}\delta\left(  E_{N}-E_{N^{\prime}}+\epsilon_{\ell}%
-\epsilon_{\ell^{^{\prime}}}\right)  $ and $\omega_{\ell^{^{\prime}}\ell
}^{\left(  2\right)  }f_{\ell}^{0}\left(  1-f_{\ell^{^{\prime}}}^{0}\right)
-\omega_{\ell\ell^{^{\prime}}}^{\left(  2\right)  }f_{\ell^{^{\prime}}}%
^{0}\left(  1-f_{\ell}^{0}\right)  =0$.

Summarizing, in the case of $A=\mathbf{v}$ we get
\begin{equation}
\delta^{\text{sj}}\mathbf{v}=\sum_{\ell\ell^{^{\prime}}}^{\prime}\left[
f_{\ell}\left(  1-f_{\ell^{^{\prime}}}\right)  \right]  ^{\left(  -2\right)
}\omega_{\ell^{^{\prime}}\ell}^{\left(  2\right)  }\delta\mathbf{r}%
_{\ell^{^{\prime}}\ell}=\sum_{\ell}f_{\ell}^{\left(  -2\right)  }\left[
\sum_{\ell^{^{\prime}}}\frac{1-f_{\ell^{^{\prime}}}^{0}}{1-f_{\ell}^{0}}%
\omega_{\ell^{^{\prime}}\ell}^{\left(  2\right)  }\delta\mathbf{r}%
_{\ell^{^{\prime}}\ell}\right]  ,
\end{equation}
where we have used Eqs. (\ref{scattering rate}) and (\ref{detailed balance})
as well as the two statistical assumptions (\ref{statistical-1}) and
(\ref{AMC}), and applied the techniques used in Appendix A of Ref.
\cite{Xiao2017SOT-SBE}. This result confirms our heuristic argument on the
\textquotedblleft proper definition\textquotedblright\ of the semiclassical
side-jump velocity $\mathbf{v}_{\ell}^{\text{sj}}=\sum_{\ell^{^{\prime}}}%
\frac{1-f_{\ell^{^{\prime}}}^{0}}{1-f_{\ell}^{0}}\omega_{\ell^{^{\prime}}\ell
}^{\left(  2\right)  }\delta\mathbf{r}_{\ell^{^{\prime}}\ell}$ in the case of
dynamical disorder in the main text (note that $\omega_{\ell^{^{\prime}}\ell
}^{\left(  2\right)  }\equiv w_{\ell^{\prime}\ell}$ and $f_{\ell}^{\left(
-2\right)  }=\delta f_{\ell}^{n}$).

Similar to the case of static disorder, the interband-coherence nature of
$\mathbf{v}_{\ell}^{\text{sj}}$ and thus that of the anomalous distribution
function $g_{\ell}^{a}$ are not quite obvious when $\mathbf{v}_{\ell
}^{\text{sj}}$ is expressed in terms of $\delta\mathbf{r}_{\ell^{^{\prime}%
}\ell}$ \cite{Xiao2017SOT-SBE,Xiao2018KL}. Therefore, in the following we
provide some more information about scattering-induced interband-coherence
response $\delta^{\text{sj}}A$ when $A$ is not necessarily the current
\cite{Xiao2017SOT-SBE,Xiao2018KL}. In the following derivation the
interband-coherence nature of $\mathbf{v}_{\ell}^{\text{sj}}$ is apparent. In
general cases of $A$, we have
\begin{align*}
&  \sum_{nN,n^{\prime}N^{\prime},n^{\prime\prime}N^{\prime\prime}}^{\prime
}\tilde{F}_{nN}^{\left(  -2\right)  }\frac{\tilde{H}_{nN,n^{\prime}N^{\prime}%
}^{^{\prime}}\tilde{H}_{n^{\prime\prime}N^{\prime\prime},nN}^{^{\prime}}%
\tilde{A}_{n^{\prime},n^{\prime\prime}}}{\left(  E_{nN}-E_{n^{\prime}%
N^{\prime}}-i\hbar s\right)  \left(  E_{nN}-E_{n^{\prime\prime}N^{\prime
\prime}}+i\hbar s\right)  }\\
&  =2\operatorname{Re}\sum_{\ell jj^{\prime}}^{\prime}\sum_{n}\sum
_{N,N^{\prime}}\tilde{F}_{nN}^{\left(  -2\right)  }n_{\ell}\left(
1-n_{j}\right)  \frac{H_{\ell N,jN^{\prime}}^{^{\prime}}A_{jj^{\prime}%
}H_{j^{\prime}N^{\prime},\ell N}^{^{\prime}}}{\epsilon_{j}-\epsilon
_{j^{\prime}}}\frac{1}{E_{N}-E_{N^{\prime}}+\epsilon_{\ell}-\epsilon
_{j}-i\hbar s}\\
&  -2\operatorname{Re}\sum_{\ell jj^{\prime}}^{\prime}\sum_{n}\sum
_{N,N^{\prime}}\tilde{F}_{nN}^{\left(  -2\right)  }n_{\ell}\left(
1-n_{j}\right)  n_{j^{\prime}}\frac{H_{\ell N,jN^{\prime}}^{^{\prime}%
}A_{jj^{\prime}}H_{j^{\prime}N^{\prime},\ell N}^{^{\prime}}}{\epsilon
_{j}-\epsilon_{j^{\prime}}}\frac{1}{E_{N}-E_{N^{\prime}}+\epsilon_{\ell
}-\epsilon_{j}-i\hbar s}\\
&  +2\operatorname{Re}\sum_{\ell jj^{\prime}}^{\prime}\sum_{n}\sum
_{N,N^{\prime}}\tilde{F}_{nN}^{\left(  -2\right)  }\left(  1-n_{j}\right)
n_{\ell}n_{j^{\prime}}\frac{H_{\ell N,jN^{\prime}}^{^{\prime}}H_{jN^{\prime
},j^{\prime}N}^{^{\prime}}A_{j^{\prime}\ell}}{\epsilon_{\ell}-\epsilon
_{j^{\prime}}}\frac{1}{E_{N}-E_{N^{\prime}}+\epsilon_{\ell}-\epsilon
_{j}-i\hbar s}%
\end{align*}
and%
\begin{align*}
&  \sum_{nN,n^{\prime}N^{\prime},n^{\prime\prime}N^{\prime\prime}}^{\prime
}\tilde{F}_{nN}^{\left(  -2\right)  }\left[  \frac{\tilde{H}_{n^{\prime
}N^{\prime},n^{\prime\prime}N^{\prime\prime}}^{^{\prime}}\tilde{H}%
_{n^{\prime\prime}N^{\prime\prime},nN}^{^{\prime}}\tilde{A}_{n,n^{\prime}}%
}{\left(  E_{nN}-E_{n^{\prime}N^{\prime}}+i\hbar s\right)  \left(
E_{nN}-E_{n^{\prime\prime}N^{\prime\prime}}+i\hbar s\right)  }+c.c.\right]  \\
&  =2\operatorname{Re}\sum_{\ell jj^{\prime}}^{\prime}\sum_{n}\sum
_{N,N^{\prime}}\tilde{F}_{nN}^{\left(  -2\right)  }n_{\ell}\left(
1-n_{j}\right)  \left(  1-n_{j^{\prime}}\right)  \frac{H_{\ell N,jN^{\prime}%
}^{^{\prime}}H_{jN^{\prime},j^{\prime}N}^{^{\prime}}A_{j^{\prime}\ell}%
}{\left(  \epsilon_{\ell}-\epsilon_{j^{\prime}}-i\hbar s\right)  \left(
E_{N}-E_{N^{\prime}}+\epsilon_{\ell}-\epsilon_{j}-i\hbar s\right)  }\\
&  -2\operatorname{Re}\sum_{\ell jj^{\prime}}^{\prime}\sum_{n}\sum
_{N,N^{\prime}}\tilde{F}_{nN}^{\left(  -2\right)  }\frac{H_{\ell N,jN^{\prime
}}^{^{\prime}}A_{jj^{\prime}}H_{j^{\prime}N^{\prime},\ell N}^{^{\prime}%
}n_{\ell}\left(  1-n_{j}\right)  n_{j^{\prime}}}{\left(  \epsilon_{j^{\prime}%
}-\epsilon_{j}-i\hbar s\right)  \left(  E_{N}-E_{N^{\prime}}+\epsilon_{\ell
}-\epsilon_{j}-i\hbar s\right)  },
\end{align*}
thus by some permutation of indices we get%
\begin{align}
&  \delta^{\text{sj}}A=2\operatorname{Re}\sum_{\ell jj^{\prime}}^{\prime}%
\sum_{n}\sum_{N,N^{\prime}}\tilde{F}_{nN}^{\left(  -2\right)  }n_{\ell}\left(
1-n_{j}\right)  \frac{1}{E_{N}-E_{N^{\prime}}+\epsilon_{\ell}-\epsilon
_{j}-i\hbar s}H_{\ell N,jN^{\prime}}^{^{\prime}}\left[  \frac{A_{jj^{\prime}%
}H_{j^{\prime}N^{\prime},\ell N}^{^{\prime}}}{\epsilon_{j}-\epsilon
_{j^{\prime}}}+\frac{H_{jN^{\prime},j^{\prime}N}^{^{\prime}}A_{j^{\prime}\ell
}}{\epsilon_{\ell}-\epsilon_{j^{\prime}}}\right]  \nonumber\\
&  =2\operatorname{Re}\sum_{\ell\ell^{^{\prime}}j^{\prime}}^{\prime}f_{\ell
}^{\left(  -2\right)  }\left[  \left(  1-f_{\ell^{^{\prime}}}^{0}\right)
\sum_{N,N^{\prime}}P_{N}^{\left(  0\right)  }+f_{\ell^{^{\prime}}}^{0}%
\sum_{N,N^{\prime}}P_{N^{\prime}}^{\left(  0\right)  }\right]  \frac{H_{\ell
N,\ell^{^{\prime}}N^{\prime}}^{^{\prime}}}{E_{N}-E_{N^{\prime}}+\epsilon
_{\ell}-\epsilon_{\ell^{^{\prime}}}-i\hbar s}\left[  \frac{H_{\ell^{^{\prime}%
}N^{\prime},j^{\prime}N}^{^{\prime}}A_{j^{\prime}\ell}}{\epsilon_{\ell
}-\epsilon_{j^{\prime}}}-\frac{A_{\ell^{^{\prime}}j^{\prime}}H_{j^{\prime
}N^{\prime},\ell N}^{^{\prime}}}{\epsilon_{j^{\prime}}-\epsilon_{\ell
^{^{\prime}}}}\right]  ,
\end{align}
i.e., $\delta^{\text{sj}}A=\sum_{\ell}f_{\ell}^{\left(  -2\right)  }%
\delta^{\text{sj}}A_{\ell}$ with%
\begin{equation}
\delta^{\text{sj}}A_{\ell}=2\operatorname{Re}\sum_{\ell^{^{\prime}}j^{\prime}%
}^{\prime}\left[  \left(  1-f_{\ell^{^{\prime}}}^{0}\right)  \sum
_{N,N^{\prime}}P_{N}^{\left(  0\right)  }+f_{\ell^{^{\prime}}}^{0}%
\sum_{N,N^{\prime}}P_{N^{\prime}}^{\left(  0\right)  }\right]  \frac{H_{\ell
N,\ell^{^{\prime}}N^{\prime}}^{^{\prime}}}{E_{N}-E_{N^{\prime}}+\epsilon
_{\ell}-\epsilon_{\ell^{^{\prime}}}-i\hbar s}\left[  \frac{H_{\ell^{^{\prime}%
}N^{\prime},j^{\prime}N}^{^{\prime}}A_{j^{\prime}\ell}}{\epsilon_{\ell
}-\epsilon_{j^{\prime}}}-\frac{A_{\ell^{^{\prime}}j^{\prime}}H_{j^{\prime
}N^{\prime},\ell N}^{^{\prime}}}{\epsilon_{j^{\prime}}-\epsilon_{\ell
^{^{\prime}}}}\right]  .
\end{equation}
From the interband matrix elements $A_{jj^{\prime}}$ and $A_{j^{\prime}\ell}$
(the momenta of the two states denoted by the subscripts are equal) one can
see that the interband-coherence plays a role in both terms.

For static impurities, the state of the scattering system remains unchanged
thus $N=N^{^{\prime}}$, and
\begin{equation}
\sum_{N,N^{\prime}}P_{N}^{\left(  0\right)  }H_{\ell N,\ell^{^{\prime}%
}N^{\prime}}^{^{\prime}}H_{\ell^{^{\prime}}N^{\prime},j^{\prime}N}^{^{\prime}%
}=\sum_{N}P_{N}^{\left(  0\right)  }H_{\ell N,\ell^{^{\prime}}N}^{^{\prime}%
}H_{\ell^{^{\prime}}N,j^{\prime}N}^{^{\prime}}=\left\langle H_{\ell
\ell^{^{\prime}}}^{^{\prime}}H_{\ell^{^{\prime}}j^{\prime}}^{^{\prime}%
}\right\rangle
\end{equation}
is just the average over the disorder configurations. Therefore, after some
algebra we obtain
\begin{equation}
\delta^{\text{sj}}A=\sum_{\ell}f_{\ell}^{\left(  -2\right)  }\left[
\sum_{\ell^{^{\prime}},\ell^{\prime\prime}\neq\ell^{^{\prime}}}\frac
{\left\langle H_{\ell\ell^{\prime}}^{^{\prime}}H_{\ell^{\prime\prime}\ell
}^{^{\prime}}\right\rangle A_{\ell^{\prime}\ell^{\prime\prime}}}{\left(
\epsilon_{\ell}-\epsilon_{\ell^{\prime}}-i\hbar s\right)  \left(
\epsilon_{\ell}-\epsilon_{\ell^{\prime\prime}}+i\hbar s\right)  }%
+2\operatorname{Re}\sum_{\ell^{\prime}\neq\ell,\ell^{\prime\prime}}%
\frac{\left\langle H_{\ell^{\prime}\ell^{\prime\prime}}^{^{\prime}}%
H_{\ell^{\prime\prime}\ell}^{^{\prime}}\right\rangle A_{\ell\ell^{\prime}}%
}{\left(  \epsilon_{\ell}-\epsilon_{\ell^{\prime}}+i\hbar s\right)  \left(
\epsilon_{\ell}-\epsilon_{\ell^{\prime\prime}}+i\hbar s\right)  }\right]  ,
\end{equation}
which just reproduces the result obtained in the single-particle T-matrix
formalism in the case of static disorder \cite{Xiao2017SOT-SBE,Xiao2018KL}.

\section{Generalized Bloch-Boltzmann formalism from the Lyo-Holstein transport theory}

The Lyo-Holstein theory \cite{Lyo1973,Holstein1964} takes into account the
many-body effects in weakly-coupled electron-phonon systems. Lyo
\cite{Lyo1973} split the electron coordinate operator into intra-cell and
inter-cell parts and considered separately the resulting four components of
the velocity-velocity correlation function. The theory thus contains some non-gauge-invariant quantities which are difficult to interpret. Partly because of these
complications, the theory has not found wide applications. The main theoretical
results of Lyo are his Eqs. (3.39) and (3.43). The latter representing the
crossed part of intrinsic skew scattering appears in the third Born order and
is too complicated to be applicable in practice. We focus on Lyo's Eq. (3.39),
which contains the contents of Lyo's Eqs. (3.25) -- (3.27), (3.37) and (3.38).
We show that, Lyo's Eq. (3.39) includes the intrinsic and side jump
anomalous Hall conductivities. The proof of the equivalence are outlined as
the following four steps:

$\left(  \text{I}\right)  $ Lyo's transport equation (3.27) is our Eq. (2b) in
the main text for $g_{\ell}^{n}$, i.e., the conventional Bloch-Boltzmann equation.

$\left(  \text{II}\right)  $ The opposite of the anomalous velocity defined by
Lyo's Eq. (3.26) is the last term of our side-jump velocity:
\begin{equation}
\mathbf{v}_{\ell}^{\text{sj},Lyo}=\sum_{\ell^{^{\prime}}}w_{\ell^{^{\prime}%
}\ell}\frac{1-f_{\ell^{^{\prime}}}^{0}}{1-f_{\ell}^{0}}\left(  -\mathbf{\hat
{D}}\arg V_{\ell^{^{\prime}}\ell}\right)  .
\end{equation}
Here $w_{\ell^{^{\prime}}\ell}$\ is the electron-phonon scattering rate taking
the same form as the lowest-Born-order expression in the density matrix
approach, but with all the quantities renormalized by many-body effects
(RPA-type renormalizations). For example, $w_{\ell^{^{\prime}}\ell}^{\left(
2\right)  }$ is proportional to $\left\vert V_{\ell\ell^{\prime}}\right\vert
^{2}$ with the renormalized electron-phonon coupling $V_{\ell\ell^{\prime}}$.
But Lyo's anomalous velocity is not gauge invariant (under the gauge
transformation $|u_{\ell}\rangle\rightarrow e^{i\theta_{\ell}}|u_{\ell}%
\rangle$).

$\left(  \text{III}\right)  $ Lyo's transport equation (3.37) corresponds to
our Eq. (11) in the main text for the anomalous distribution function
$g_{\ell}^{a}$, but has a different form
\begin{equation}
e\mathbf{E}\cdot\mathbf{v}_{\ell}^{\text{sj},Lyo}=-\sum_{\ell^{^{\prime}}%
}w_{\ell^{^{\prime}}\ell}\frac{1-f_{\ell^{^{\prime}}}^{0}}{1-f_{\ell}^{0}%
}\left(  g_{\ell}^{a,Lyo}-g_{\ell^{^{\prime}}}^{a,Lyo}\right)  ,
\end{equation}
because Lyo defined his transport function as
\begin{equation}
g_{\ell}^{a,Lyo}=g_{\ell}^{a}-e\mathbf{E}\cdot\mathbf{A}_{\ell},
\end{equation}
with $\mathbf{A}_{\ell}$ the Berry connection. The so-defined transport
function is not gauge invariant and not a real distribution function.

$\left(  \text{IV}\right)  $ Combining $\left(  \text{I}\right)  -\left(
\text{III}\right)  $, we recognize that Lyo's Eqs. (3.25) and (3.38), whose
sum gives his (3.39), take the following form in our notations:
\begin{equation}
\left(  j^{e}\right)  _{y}^{Lyo-\text{sj(1)}}=e\sum_{\ell}\left(
-\frac{\partial f_{\ell}^{0}}{\partial\epsilon_{\ell}}\right)  g_{\ell}%
^{n}\left(  \mathbf{v}_{\ell}^{\text{sj},Lyo}\right)  _{y},\label{Lyo-sj}%
\end{equation}%
\begin{equation}
\left(  j^{e}\right)  _{y}^{Lyo-\text{sj(2)}}=e\sum_{\ell}\left(
-\frac{\partial f_{\ell}^{0}}{\partial\epsilon_{\ell}}\right)  g_{\ell
}^{a,Lyo}\left(  \mathbf{v}_{\ell}^{0}\right)  _{y}.\label{Lyo-a}%
\end{equation}
Both of them are gauge dependent. But we show that the sum of them is gauge
invariant. In fact we show
\begin{equation}
\left(  j^{e}\right)  _{y}^{\text{sj(1)}}=\left(  j^{e}\right)  _{y}%
^{Lyo-\text{sj(1)}}-e^{2}E_{x}\sum_{\ell}\left(  -\frac{\partial f_{\ell}^{0}%
}{\partial\epsilon_{\ell}}\right)  \left(  A_{\ell}\right)  _{y}\left(
v_{\ell}^{0}\right)  _{x},\label{link}%
\end{equation}
and%
\begin{equation}
\left(  j^{e}\right)  _{y}^{\text{sj(2)}}=\left(  j^{e}\right)  _{y}%
^{Lyo-\text{sj(2)}}+e^{2}E_{x}\sum_{\ell}\left(  -\frac{\partial f_{\ell}^{0}%
}{\partial\epsilon_{\ell}}\right)  \left(  A_{\ell}\right)  _{x}\left(
v_{\ell}^{0}\right)  _{y},
\end{equation}
thus
\begin{equation}
\left(  j^{e}\right)  _{y}^{Lyo-\text{sj(1)}}+\left(  j^{e}\right)
_{y}^{Lyo-\text{sj(2)}}=\left(  j^{e}\right)  _{y}^{\text{sj(1)}}+\left(
j^{e}\right)  _{y}^{\text{sj(2)}}+\left(  j^{e}\right)  _{y}^{\text{in}}.
\end{equation}
As an example we provide the derivation of Eq. (\ref{link}):
\begin{align*}
&  \left(  j^{e}\right)  _{y}^{\text{sj(1)}}-\left(  j^{e}\right)
_{y}^{Lyo-\text{sj(1)}}\\
&  =e\sum_{\ell,\ell^{^{\prime}}}\left(  -\frac{\partial f^{0}}{\partial
\epsilon_{\ell}}\right)  g_{\ell}^{n}w_{\ell^{^{\prime}}\ell}\frac
{1-f^{0}\left(  \epsilon_{\ell^{^{\prime}}}\right)  }{1-f^{0}\left(
\epsilon_{\ell}\right)  }\left[  -\left(  A_{\ell}\right)  _{y}\right]
+e\sum_{\ell,\ell^{^{\prime}}}\delta f_{\ell^{^{\prime}}}w_{\ell\ell
^{^{\prime}}}\frac{1-f^{0}\left(  \epsilon_{\ell}\right)  }{1-f^{0}\left(
\epsilon_{\ell^{^{\prime}}}\right)  }\left(  A_{\ell}\right)  _{y}\\
%&  =e\sum_{\ell,\ell^{^{\prime}}}\left(  -\frac{\partial f^{0}}{\partial
%\epsilon_{\ell}}\right)  \left(  A_{\ell}\right)  _{y}w_{\ell^{^{\prime}}\ell
%}\frac{1-f^{0}\left(  \epsilon_{\ell^{^{\prime}}}\right)  }{1-f^{0}\left(
%\epsilon_{\ell}\right)  }\left[  -g_{\ell}^{n}\right]  +e\sum_{\ell
%,\ell^{^{\prime}}}\left(  A_{\ell}\right)  _{y}\left(  -\frac{\partial f^{0}%
%}{\partial\epsilon_{\ell^{^{\prime}}}}\right)  g_{\ell^{^{\prime}}}^{n}%
%w_{\ell^{^{\prime}}\ell}e^{\beta\left(  \epsilon_{\ell^{^{\prime}}}%
%-\epsilon_{\ell}\right)  }\frac{1-f^{0}\left(  \epsilon_{\ell}\right)
%}{1-f^{0}\left(  \epsilon_{\ell^{^{\prime}}}\right)  }\\
&  =e\sum_{\ell}\left(  -\frac{\partial f^{0}}{\partial\epsilon_{\ell}%
}\right)  \left(  A_{\ell}\right)  _{y}\sum_{\ell^{^{\prime}}}w_{\ell
^{^{\prime}}\ell}\frac{1-f^{0}\left(  \epsilon_{\ell^{^{\prime}}}\right)
}{1-f^{0}\left(  \epsilon_{\ell}\right)  }\left[  g_{\ell^{^{\prime}}}%
^{n}-g_{\ell}^{n}\right]  \\
&  =-e^{2}E_{x}\sum_{\ell}\left(  -\frac{\partial f^{0}}{\partial
\epsilon_{\ell}}\right)  \left(  A_{\ell}\right)  _{y}\left(  v_{\ell}%
^{0}\right)  _{x},
\end{align*}
where the interchange of $\ell$ and $\ell^{^{\prime}}$ is used in the first
step and the conventional Bloch-Boltzmann equation of the main text is
used in the last step.
\end{widetext}

\end{document}